\documentclass[12pt]{iopart}
\usepackage{iopams}
\usepackage{graphicx}
\usepackage{epsfig}
\usepackage{cite}
\usepackage{epstopdf}

\begin{document}

\maketitle
\title{Cooperator driven oscillation in a time-delayed feedback-evolving game}
\author{Fang Yan$^{1,2}$, Xiaojie Chen$^1$,
Zhipeng Qiu$^3$  and Attila Szolnoki$^4$}

\address{$^1$ School of Mathematical Sciences, University of
Electronic Science and Technology of China, Chengdu 611731,
 People's Republic of
China}
\address{$^2$ School of Automation Engineering, University of
Electronic Science and Technology of China, Chengdu 611731, People's Republic of
China}
\address{$^3$ Department of Mathematics, Nanjing University of Science and Technology, Nanjing 210094, People's Republic of
China}
\address{$^4$ Institute of Technical Physics and Materials Science, Centre for Energy Research, P.O. Box 49, Budapest H-1525, Hungary}
\ead{xiaojiechen@uestc.edu.cn}
\vspace{10pt}
%\begin{indented}
%\item[]January 2021
%\end{indented}

\begin{abstract}
\\
Considering feedback of collective actions of cooperation on common resources has vital importance to reach sustainability. But such efforts may have not immediate consequence on the state of environment and it is unclear how they influence the strategic and environmental dynamics with feedbacks. To address this issue, we construct a feedback-evolving game model in which we consider the growth capacity of resources and the punishment efficiency on defectors who do not provide returns to the environment. Importantly, we further assume a delay in adopting the contribution of cooperative individuals to environmental change in our model. We find that when this contribution amount from cooperators' endowment is fixed, the time delay has no particular consequence on the coevolutionary dynamics. However, when the return is proportional to their endowment, then the time delay can induce periodic oscillatory dynamics of cooperation level and environment. Our work reveals the potential effects of time delay of cooperative actions on the coevolutionary dynamics in strategic interactions with environmental feedback.

\end{abstract}

\vspace{2pc}
\noindent{\it Keywords}: cooperation, common-pool resource, feedback-evolving game, time-delay, oscillation

%\maketitle

\section{Introduction}

The sustainable use of common-pool resources
depends crucially on the interdependence of resource and social dynamics \cite{ostrom_90,brander_aer98,hauser_n14,estrela_tee19,sugiarto_prl17}. Indeed, there is an environmental feedback between available resources and the strategies of users: an individual's payoff relies not exclusively on other's action, but also on the actual state of the resources. Furthermore, the latter is also influenced by the actions of individuals forming the population. Because of its importance, the mentioned feedback becomes a decisive component of environmental modeling for studying the governance of common-pool resources in recent years \cite{santos_pnas11,sanchez_pb13,allen_pb13,vasconcelos_pnas14,pacheco_plrev14,tavoni_ncc14,hilbe_n18,su_pnas19,barfuss_pnas20}.

The subtle interdependence of resource and social dynamics can be grasped via feedback-evolving game models, which have attracted intensive research activity in recent years \cite{tavoni_jtb12,lade_te13,weitz_pnas16,lee_jh_jtb17,chen_xj_pcb18,shao_yx_epl19,hauert_jtb19,lin_yh_prl19,wang_x_rspa20,tilman_nc20}. For example, Weitz \emph{et al.} observed oscillations of strategies and the environment in a feedback-evolving game model \cite{weitz_pnas16}, and similar periodic state was reported in asymmetric games due to environmental heterogeneity \cite{hauert_jtb19}. In the framework of feedback-evolving game, some works have further demonstrated that the governance of the commons can be controlled by institutions \cite{sigmund_n10,vasconcelos_ncc13,han_jrsif15,perc_pr17}. It is proved that introducing ostracism can maintain cooperation in resource usage under variable social and environmental conditions \cite{tavoni_jtb12}. Albeit delicately adjusted punishment is fundamental, it is shown that the punishment effects on the governance of the commons also depend on the growing capacity of renewable resources \cite{chen_xj_pcb18}.

The mentioned coevolutionary models, however, have skipped an important feature of feedback mechanisms, which is a potential time-delay of individual actions on the governance of the common resources. Notably, the latter could be a decisive feature in case of renewable resources, when it takes a while to manifest improvements after a positive action. For example, the consequence of the changes of fishermen's attitude can be realized in the improvements of fisher stocks after a while \cite{kraak_ff11}. In addition, negative acts like polluting soil might have consequence only in the next year's harvest. Accordingly, such time-delay feature is different from the relative timescale of strategy and resource dynamics introduced in previous works~\cite{weitz_pnas16,tilman_nc20}, which characterizes the relative strength of strategic versus resource change. On the other hand, we stress that delays in fitness adjustment have been already considered by Bauer and Frey who observed a coexistence of two competing species in a metapopulation \cite{bauer_prl18}. But studying the direct consequence of time delay on a renewable environment which is subject to the battle of competing consuming strategies remained unexplored.

To clarify the potential consequence of time delay on environmental change, here we propose a feedback-evolving game where cooperators and defectors compete for common resources.  While both defectors and cooperators are allocated with the same amount from the common pool, but cooperators reinvest a certain amount back to the environment to maintain sustainability. Defectors, who do nothing for this purpose, are monitored and punished institutionally with a certain probability. The key question is how to adopt the cooperators' contributions to environmental change. Here we assume a certain time delay in adopting the contribution of cooperators to environmental change. For a comprehensive understanding, we consider two significantly different scenarios regarding how cooperators make contributions to the common pool, which practically covers realistic options \cite{santos_nature08,szolnoki_njp16,rauch_jrsif17,chen_xj_pcb18}. In the first case, the contribution amount from cooperators is fixed and independent of their endowment from the common resource. The second option considers the fact how intensively the environment is utilized by the consumers. In this case, the applied reinvestment is proportional to their dynamical endowment from the common resource.

By means of theoretical analysis and numerical calculations, we find that there is a conceptual difference in the system's behavior depending on how the reinvestment of cooperators is applied to environmental recovery. When cooperators make a fixed contribution to the common pool, the evolutionary outcome is insensitive to the applied time delay. On the other hand, when the amount of cooperators to the common pool is proportional to their endowment, the introduction of time delay can induce periodic oscillations of cooperation level and resources. More precisely, there exists a critical time delay at which a Hopf bifurcation occurs. Furthermore, we can determine the direction of Hopf bifurcation and the stability of the bifurcating periodic solutions by using normal form theory and center manifold theorem~\cite{hassard_81}.

\section{Model and Methods}

We consider a population of size $N$ where two basic consuming strategies, i.e., cooperation and defection, compete for common-pool resources. While the time-dependent resource amount $y(t)$ is limited, but it is partly renewable and its dynamics can be described by the well-known logistic population growth model~\cite{tsoularis_mb02}, given by $\dot{y}(t)=ry(t)[1-\frac{y(t)}{R_{\rm m}}]$, where $r$ is the intrinsic growth rate and $R_{\rm m}$ is the carrying capacity of resource pool. Meanwhile, each individual can receive an initial endowment from the common pool, which represents the harvesting amount from the common pool and is given by $\frac{y(t)}{R_{\rm m}}b_{\rm m}$, where $b_{\rm m}$ is the maximal resource portion that each individual is capable to obtain per unit of time when the amount of the common pool resource $y(t)$ reaches the carrying capacity of resource pool $R_{\rm m}$. To implement the difference in consuming attitudes, we further assume that cooperators reinvest a certain amount back to the common pool to prevent depletion. Defectors, however, do nothing for this purpose. Based on previous observations \cite{tavoni_jtb12,yang_pnas13,chen_njp14,chen_srep15,liu_mmmas19}, we assume that consumers are monitored and defection is punished by a centrally organized management. It is detected with a probability $p$ ($0<p<1$) and the involved defector is punished with a fine $\beta$ ($\beta>0$) which is deducted from the individual's payoff.

\begin{figure*}
   \centering
   \fbox{
     \begin{minipage}{6.0 in}
       \centering
       \includegraphics[width=4in]{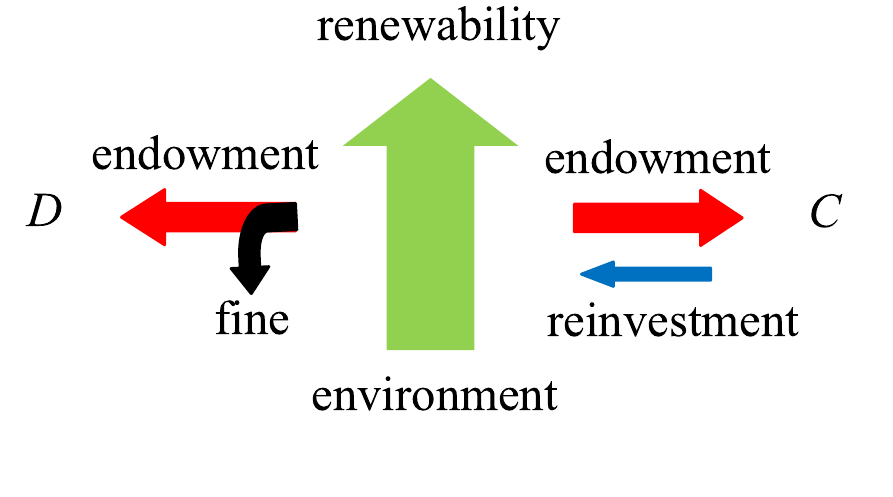}
       \caption{Blueprint of coevolutionary dynamics of strategies and environment. The latter's state would change via a logistic growth of intrinsic dynamics, but both defectors and cooperators utilize resources. While defection may be identified and punished, cooperators reinvest an amount to the common pool. Importantly, its consequence on the environment manifests only after some delay. For the dynamical process the key point is whether this amount is fixed or proportional to the cooperator's endowment.}
       \label{model}
     \end{minipage}}
\end{figure*}

The key elements of the proposed model are summarized in figure~\ref{model}. Here green arrow showing up represents the fact that the environment has an intrinsic dynamic feature with a renewable capacity. Technically, this is described by the logistic population growth part. Consumers, independently whether they are defectors or cooperators, enjoy the available resources which are signed by red arrows. Cooperators are responsible for avoiding resource depletion, therefore they invest back an amount to the environment. This act is marked by a blue arrow. It is important to stress, however, that the consequence of this investment to the environment's state can be realized only after a time delay $\tau$. Last we note that defectors, who do not bother with the state of the environment, may be punished and the fine is deducted from the related payoff value.

We consider a finite, but large well-mixed population and use the replicator equation to describe the time evolution of cooperation level \cite{hofbauer_98,sandholm_2011,tanimoto_2015}. Accordingly, we have
\begin{eqnarray}
\dot{x}(t)=x(t)(1-x(t))[P_C(t)-P_D(t)], \nonumber
\end{eqnarray}
where $x(t)$ is the fraction of cooperators in the population at time $t$, while $P_C(t)$ and $P_D(t)$ are the payoff values of cooperators and defectors at time $t$, respectively. We emphasize that the replicator equation is often used to study the strategic dynamics in infinite well-mixed populations \cite{hofbauer_98}, but the classical stability theory of the replicator dynamics is still valid in the large finite population limit \cite{harper_entropy16}. This is because for large populations the fluctuations in the fraction of individuals of a given strategy induced by stochastic noise become increasingly small compared to their actual value, and we can find that the ordinary differential equation of the system neglecting the stochastic term are closely related to the replicator equation \cite{pacheco_plrev14}.

In the first case, we assume that cooperators invest a fixed amount of $g$ to the common pool, hence the mentioned payoff values can be written as $P_{C}(t) = \frac{b_{\rm m}y(t)}{R_{\rm m}}-g$ for cooperators and $P_{D}(t) = \frac{b_{\rm m}y(t)}{R_{\rm m}} - p\beta$ for defectors. Importantly, the reinvestment to environment is considered with a time delay, hence the proper equation system for cooperation level and environment is
\begin{equation}
\left\{
\begin{array}{ll}
\dot{x}(t)=x(t)[1-x(t)](p\beta-g)\\
\dot{y}(t)=ry(t)[1-\frac{y(t)}{R_{\rm m}}]-N\frac{y(t)}{R_{\rm m}}b_{\rm m}+gNx(t-\tau)\,\,.
\end{array}
\right.
\label{fix}
\end{equation}
By solving these equations, we find that the evolutionary outcome of the system behavior is irrelevant with the time delay. There is no particular consequence of how long delay is applied, and the
system states including the actual cooperation level evolve practically onto the same stationary state, which only depends on the other parameters of the model, such as $r$, $p$, or $\beta$. In appendix A, we provide theoretical analysis and numerical results for this variant of the model.

In the other case, however, we assume that cooperators reinvest an $\alpha$ ($0<\alpha\leq 1$) portion of their endowment to the environment, hence their new payoff value is $P_{C}(t) = \frac{b_{\rm m}y(t)(1-\alpha)}{R_{\rm m}}$, while a defector's payoff is not changed. Accordingly, the dynamical equations for the coupled resource-strategy system can be written as
\begin{equation}
\left\{
\begin{array}{ll}
\dot{x}(t)=x(t)[1-x(t)][p\beta-\frac{\alpha b_my(t)}{R_m}]\\
\dot{y}(t)=ry(t)[1-\frac{y(t)}{R_{\rm m}}]-N\frac{y(t)}{R_{\rm m}}b_{\rm m}+\frac{\alpha Nb_{\rm m}x(t-\tau)y(t-\tau)}{R_{\rm m}}\,\,.\label{prop}
\end{array}
\right.
\end{equation}

Let us note that the fixed points in the system depicted by equation (\ref{prop}) should be the same to those of the equation system without time delay (i.e., $\tau=0$), therefore we can obtain that
this equation system has at most five fixed points
which are $(0,0)$, $(1,0)$, $(0,R_{\rm m}-\frac{Nb_{\rm m}}{r})$,
$(1,R_{\rm m}-\frac{Nb_{\rm m}(1-\alpha)}{r})$, and
$(K,\frac{R_{\rm m}p\beta}{\alpha b_{\rm m}})$, respectively,
where $K=\frac{1}{\alpha}-\frac{rR_{\rm m}}{\alpha b_{\rm m}N}+\frac{p\beta R_{\rm m}r}{N\alpha^{2}b_{\rm m}^{2}}$. For simplicity, we use $F_0$, $F_1$, $F_2$, $F_3$, and $F_4$ to respectively represent these five fixed points.

To study the stability of these fixed points, we use the method of characteristic roots of delay differential equations \cite{kuang_93,gopalsamy_92}. For convenience, we introduce the notations of $e_{C}=\frac{Nb_{\rm m}(1-\alpha)}{R_{\rm m}}$ and $e_{D}=\frac{Nb_{\rm m}}{R_{\rm m}}$ to sign the net income of cooperators and defectors in the population from the common resource, respectively \cite{chen_xj_pcb18}. In the following, we present the corresponding results by distinguishing three substantially different parameter regions where the distinction is based on the actual intrinsic growth rate value of the renewable common pool resource.

\section{Results}

\subsection*{\rm{\textbf{3.1. Slowly growing resource pool}}}
\begin{figure}
\centering
\fbox{ \begin{minipage}{6.0 in}
\centering
\includegraphics[width=4in]{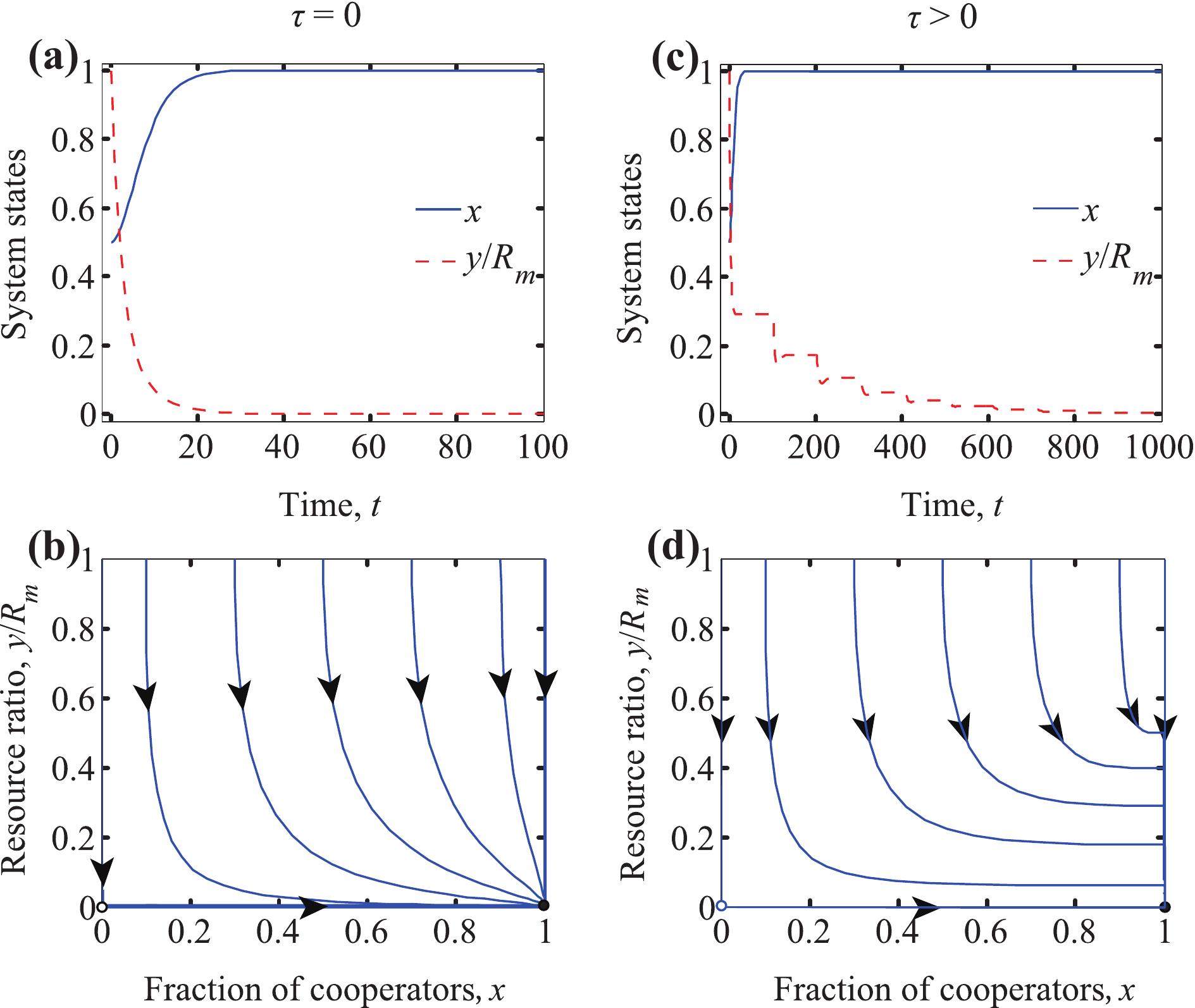}
\caption{Coevolutionary dynamics for $r < e_{C} $. Top panels show the time evolution of cooperation level and the status of resource. Bottom panels show the related phase portraits on $x - y/R_{\rm m}$ plane. Filled (open) circle represents a stable (an unstable) fixed point. Parameters are $r = 0.1$, $\alpha = 0.5$, $N = 1000$, $R_{\rm m} = 1000$, $p = 0.5$, $\beta = 0.5$, and $b_{\rm m} = 0.5$. There is no delay of feedback in the left column, while it is $\tau=100$ in the right column. Independently of the time delay, the final destination to the fixed point $F_{1}$ is inevitable.}
\label{slow}
\end{minipage}}
\end{figure}

Here the environment recovers too slowly, hence $0<r<e_{C}<e_{D}$. In this situation, the system has only two fixed points, which are $F_{0}$ and $F_{1}$, respectively. As it is discussed in appendix B, $F_{0}$ is an unstable fixed point for all $\tau\geq0$, while $F_{1}$ is asymptotically stable. In the border case of $r=e_{C}$, $F_{1}$ becomes stable, but not asymptotically stable for $\tau\geq0$. A representative time evolution of the cooperation level and the abundance of common resource is plotted in figure \ref{slow}, where we compare the cases of immediate and delayed feedbacks. It shows that the system always converges toward the fixed point $F_{1}$, no matter whether time delay is applied or not. Even if the population is driven to the full cooperator state, the resource pool becomes fully depleted, and the delayed feedback has no influence on avoiding this undesired destination.

\subsection*{\rm{\textbf{3.2. Moderately growing resource pool}}}
If the intrinsic growth rate of resources is moderate, which means $e_{C}<r<e_{D}$, the potential destinations are more subtle. Here we can distinguish two main cases in dependence of the efficiency of inspection and punishment. When the centralized institution is less effective, the term $\alpha b_{\rm m}(1-\frac{e_{C}}{r})$ exceeds $p\beta$ product. As a result, the system has four fixed points, which are $F_{0}$, $F_{1}$, $F_{3}$, and $F_{4}$, respectively. Theoretical analysis, discussed in appendix B, shows that the first three are unstable for  $\tau\geq0$, while $F_{4}$ is asymptotically stable for $\tau<\tau_{\rm c}$ and becomes unstable for $\tau>\tau_{\rm c}$. Here, we have $\tau_{\rm c}=\frac{\theta_{1}}{\omega_{+}}$,
where $\omega_{\pm}^{2}=\frac{1}{2}[{{H\pm \sqrt{H^{2}-4(c^{2}-d^{2})}}}] $ and $H=b^2+2c-a^2$, and $\theta_{1}$ satisfies
$$
\cos\theta_{1}=-\frac{(ab^{2}-d)\omega_{+}^{2}}{b^{2}\omega_{+}^{2}+d^{2}}
\hspace{0.3cm}
\textrm{and}
\hspace{0.3cm}
\sin\theta_{1}=-\frac{ad^{2}\omega_{+}+b\omega_{+}^{3}}{b^{2}\omega_{+}^{2}+d^{2}},
$$
with
$a=\frac{2rp\beta}{\alpha b_{\rm m}}+e_{D}-r$,
$b=r-\frac{rp\beta}{\alpha b_{\rm m}}-e_{D}$,
$c=0$, and
$d=Np\beta K(1-K)\frac{\alpha b_{\rm m}}{R_{\rm m}}$. In particular, for $\tau=\tau_{\rm c}$ there exists a bifurcation point where the direction of the Hopf bifurcation and the stability of the bifurcating periodic solutions can be determined.

\begin{figure*}
\centering
\fbox{ \begin{minipage}{6.0 in}
\centering
\includegraphics[width=4in]{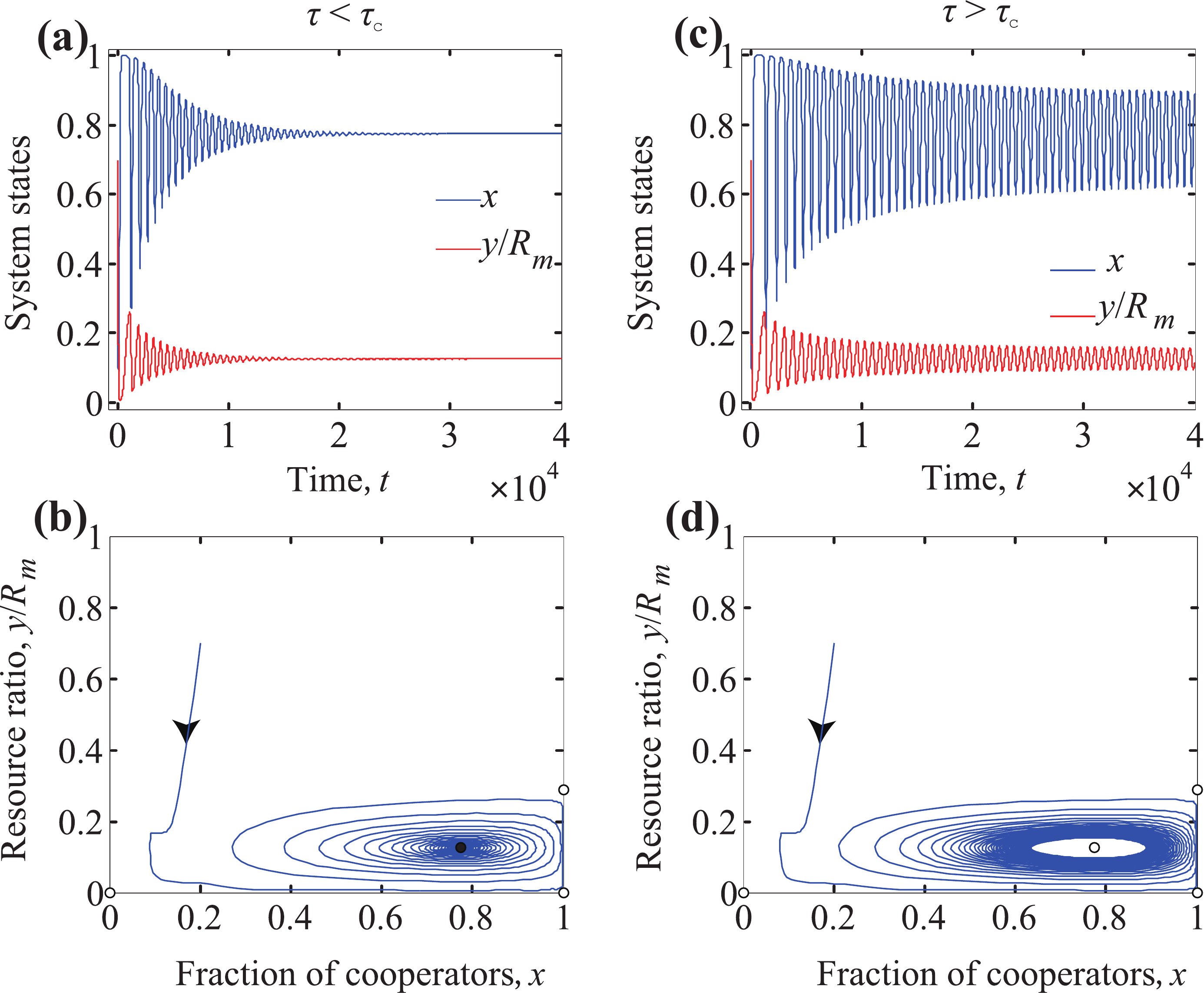}
\caption{Coevolutionary dynamics for $e_{C}<r < e_{D} $ and $0<p\beta<\alpha b_{\rm m}(1-\frac{e_{C}}{r})$. As shown in figure~\ref{slow}, filled (open) circle represents a stable (an unstable) fixed point. Parameters are $r = 0.35$, $\alpha = 0.5$, $N = 1000$, $R_{\rm m} = 1000$, $p = 0.25$, $\beta = 0.125$, and $b_{\rm m} = 0.5$. In the left column the time delay $\tau=50<\tau_{\rm c}=58.2571$ is applied and the system converges to the equilibrium point $(0.775,125)$. Right column shows the case of $\tau=59>\tau_{\rm c}$ where the system shows a persistent oscillation
of cooperation and resource.}
\label{moderate_weak}
\end{minipage}}
\end{figure*}

In figure \ref{moderate_weak}, we provide numerical examples to verify our theoretical analysis where $\tau_{\rm c}=58.2571$ for the applied parameter values. The left column shows the case when $\tau=50 <\tau_{\rm c}$ and the system converges to the fixed point $F_4$, providing a stable coexistence of cooperators and defectors at a sustainable resource level of environment. On the contrary, for $\tau=59>\tau_{\rm c}$ shown in right panels of figure \ref{moderate_weak}, the fixed point becomes unstable and the system shows persistent oscillations of cooperation level and environmental resources.  This result suggests that the magnitude of time delay can affect the coevolutionary dynamics significantly. Furthermore, as discussed in appendix B, the Hopf bifurcation occurring at $\tau_{\rm c}$ is supercritical and a stable bifurcating periodic solution emerges as $\tau$ exceeds $\tau_{\rm c}$.

The remaining case is when the environment management is effective enough, which means $p\beta>\alpha b_{\rm m}(1-\frac{e_{C}}{r})$. In this situation, the equation system has three fixed points which are $F_{0}$, $F_{1}$, and $F_{3}$, respectively. As we discuss in appendix B, the first two fixed points $F_{0}$
 and $F_{1}$ are unstable, while $F_{3}$ is asymptotically stable for $\tau\geq0$.

\begin{figure*}
\centering
\fbox{ \begin{minipage}{6.0 in}
\centering
\includegraphics[width=4in]{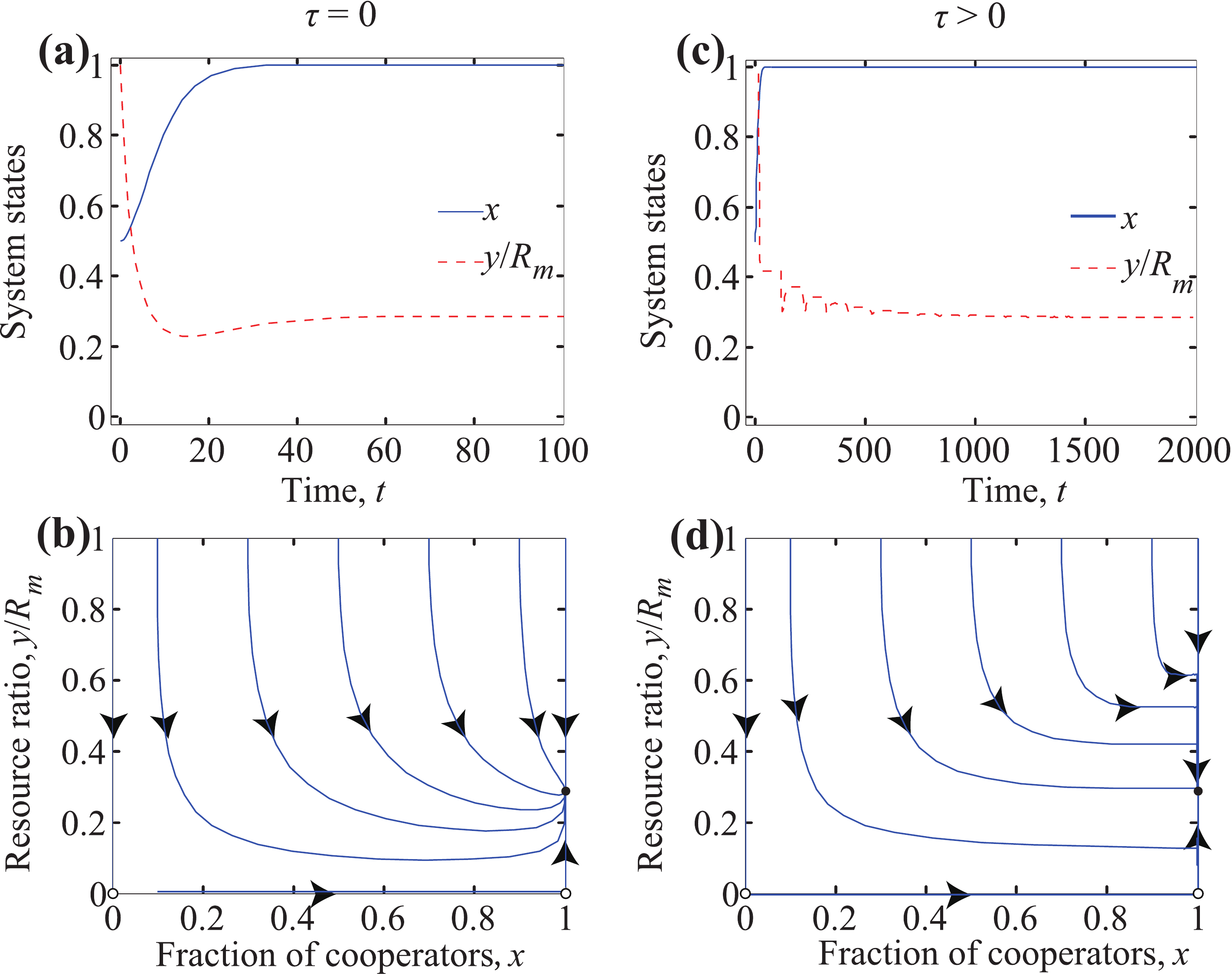}
\caption{Coevolutionary dynamics for $e_{C}<r < e_{D} $ and $p\beta>\alpha b_{\rm m}(1-\frac{e_{C}}{r})$. Notations are similar as for previous plots. Parameters are $r = 0.35$, $\alpha = 0.5$, $N = 1000$, $R_{\rm m} = 1000$, $p = 0.5$, $\beta = 0.5$, and $b_{\rm m} = 0.5$.
In the left column there is no time delay, while in the right column $\tau=100$ is applied. Here the system converges to the fixed point $(1,285.7143)$ independently of the value of $\tau$.}\label{moderate_strong}
\end{minipage}}
\end{figure*}

A representative example of the coevolutionary dynamics for this case is illustrated in figure \ref{moderate_strong}. We can see that no matter whether $\tau$ is 0 (figure \ref{moderate_strong}(a), (b)) or 100 (figure \ref{moderate_strong}(c), (d)), the  system always converges to the fixed point $F_{3}$, which is consistent with our theoretical results. This means that the stability of the equilibrium points is independent of the time delay and different from the results for slowly growing resource pool. In other words, the centralized institution has a decisive role in a sustainable resource level when the intrinsic dynamics of environment provides a necessary growth. We note that further theoretical analysis for the special border cases of $p\beta=\alpha b_{\rm m}(1-\frac{e_{C}}{r})$ and $r=e_{D}$ is presented in appendix B.

\subsection*{\rm{\textbf{3.3. Rapidly growing resource pool}}}
To explore the complete parameter space, we finally discuss the case when the intrinsic growth rate of resource is large enough to exceed $e_{D}$. According to the efficiency of inspection and punishment, we can distinguish three sub-cases here. When this institution is effective and $p\beta$ exceeds $\alpha b_{\rm m}(1-\frac{e_{C}}{r})$, then we have four fixed points, which are $F_{0}$, $F_{1}$, $F_{2}$, and $F_{3}$, respectively. Here $F_{0}$, $F_{1}$, and $F_{2}$ are unstable, while $F_{3}$ is asymptotically stable for any $\tau\geq0$ (see appendix B). The representative trajectory of evolution in this sub-case is conceptually similar to the one shown in figure \ref{moderate_strong}. It practically means that a full cooperative state can always be reached at a sustainable level of environmental resources independently of time delay.

If the above mentioned institution is less powerful, then the product $p\beta$ is less than $\alpha b_{\rm m}(1-\frac{e_{C}}{r})$, but exceeds $\alpha b_{\rm m}(1-\frac{e_{D}}{r})$. Consequently, the equation system has five fixed points which are $F_{0}$, $F_{1}$, $F_{2}$, $F_{3}$, and $F_{4}$, respectively. As proved in appendix B, the first four fixed points are unstable for any $\tau\geq0$, while $F_{4}$ is asymptotically stable for $\tau<\tau_{\rm c}$ and unstable for $\tau>\tau_{\rm c}$. For $\tau=\tau_{\rm c}$, there is a Hopf bifurcation point. Appendix B contains details of the direction and stability of bifurcation. Here the trajectories of representative evolutionary processes in this sub-case illustrate conceptually similar behavior we presented in figure \ref{moderate_weak}. More precisely, if the time delay is less than a critical value $\tau_{\rm c}$, then the system terminates onto the stable fixed point $F_4$ where cooperators and defectors coexist at a sustainable resource level. But if the time delay exceeds this critical value, then the equilibrium point $F_4$ becomes unstable and the system displays a persistent oscillatory state where the time average of cooperation level and resources are equal to the values obtained for smaller decay values. The Hopf bifurcation at $\tau_{\rm c}$ is supercritical and the bifurcating periodic solution exists when $\tau$ exceeds $\tau_{\rm c}$. Furthermore, the bifurcating periodic solution is stable. These results illustrate that the magnitude of time delay can affect the system dynamics, which are consistent with analytical predictions presented in appendix B. We can thus conclude that the mentioned institution is less powerful in this case, but it still has the ability to maintain the resource by reducing defectors for $\tau<\tau_{\rm c}$. Their fractions depend principally on the difference between resource contributions of strategies which is characterized by the parameter $\alpha$. However, for $\tau>\tau_{\rm c}$ the outcome of the coevolutionary dynamics converges to persistent oscillations of strategies and resource state.
\begin{figure*}
\centering
\fbox{ \begin{minipage}{6.0 in}
\centering
\includegraphics[width=4in]{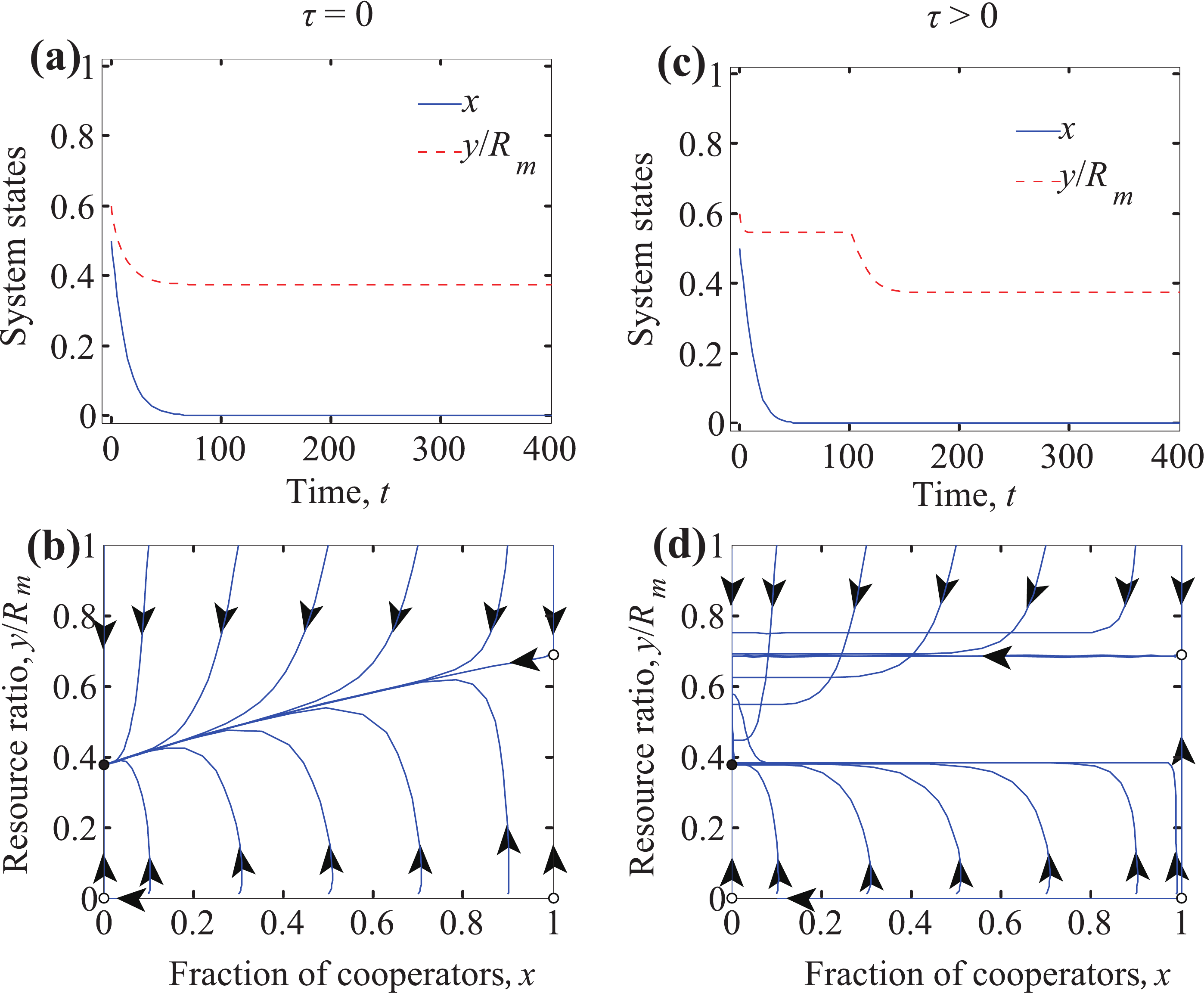}
\caption{Evolutionary trajectories for $r > e_{D} $ and $p\beta<\alpha b_{\rm m}(1-\frac{e_{D}}{r})$ when $\tau=0$ (left column) and $\tau=100$ (right column) are applied, respectively. Independently of the time delay, the system terminates onto the fixed point $(0,375)$. Parameters are $r = 0.8$, $\alpha = 0.5$, $N = 1000$, $R_{\rm m} = 1000$, $p = 0.125$, $\beta = 0.125$, and $b_{\rm m} = 0.5$.}\label{fast_weak}
\end{minipage}}
\end{figure*}

When the institution is too weak and the $p\beta$ product cannot exceed $\alpha b_{\rm m}(1-\frac{e_{D}}{r})$,
the system has four fixed points which are $F_{0}$, $F_{1}$, $F_{2}$, and $F_{3}$, respectively.
Here only $F_{2}$ is asymptotically stable for any $\tau\geq0$, while the rest are unstable for any $\tau\geq0$ (see appendix B for details). These results are illustrated in figure \ref{fast_weak} where we respectively consider $\tau=0$ (figure \ref{fast_weak}(a), (b)) and $\tau=100$ (figure \ref{fast_weak}(c), (d)). It suggests that independently of the value of $\tau$ the system terminates into the fixed point $F_{2}$. At this stable fixed point, defectors can prevail, but the strong growing capacity of environment is still capable to maintain a sustainable state.

\section{Discussion}
To investigate the long-term consequences of collective actions on the governance of common resources requires the application of feedback-evolving game models, where both individual activities and the actual state of environment coevolve in a strongly interdependent way \cite{weitz_pnas16}. Several pioneering works have realized this fact and pointed out different aspects which could be vital to control and influence the mentioned coevolution in a desired direction \cite{chen_xj_pcb18,hauert_jtb19,tilman_nc20}. There is no doubt that responsible environmental management is related with personal reinvestment into our environment. However, just a very few studies have considered the delaying effects of individual actions on the environmental change, despite of the fact that such delay is evident especially for renewable resources. In this study, we have considered such a delay factor explicitly into a feedback-evolving game model where we have also assumed a potential renewal of common resources. To distinguish personal activities, we have applied two main strategies, cooperation and defection, and assumed that cooperators are responsible for the environment and reinvest a certain portion of their endowment. This latter act is proved to be a decisive factor that may determine the coevolutionary dynamics fundamentally. More precisely, we have explored two main cases, one where the amount is fixed, the other where it is proportional to the personal harvesting amount obtained from the common pool. While the former has no particular consequence on the evolution of a delayed-feedback framework, the latter can induce significantly different system behavior.

We have shown that proportional reinvestment of individual endowment to the common resource causes the system to behave differently, in dependence on the magnitude of feedback's time delay. When the natural intrinsic growth rate of resources is not too slow and the enforcement strength is not too effective, then a Hopf bifurcation emerges as the magnitude of time delay exceeds a critical value. Beyond this, there is a persistent oscillation of cooperation and resource. Similar oscillation has been already reported by earlier works \cite{weitz_pnas16,lin_yh_prl19,sigdel_jtb17,antonioni_pre19}, but in our model there was no need to assume a two-state model to observe it. Instead, the way of cooperator's reinvestment and the magnitude of time delay are identified as the crucial factors. We note that such a high magnitude oscillation could be dangerous especially in a small system, because in the presence of noise it can easily result in an extinction \cite{dobramysl_jpa18,zheng_xd_prl18,avelino_pre18,intoy_pre15}.

Our results highlight that there is a subtle interdependence between the internal growing capacity of renewable resource, the time delay of feedback, and the environment management. They altogether determine the evolutionary outcome of such coupled strategy-resource system, and this observation should make us careful when designing any human intervention for a sustainable environment.

\section*{Acknowledgments}
This research was supported by the National Natural Science
Foundation of China (Grants Nos. 61976048 and 62036002) and the Fundamental
Research Funds of the Central Universities of China.

\appendix

\section*{Appendix A. Feedback-evolving game with time delay by using fixed contribution}
\setcounter{section}{1}
We first consider the case where cooperators contribute a fixed endowment $g$ to the common pool. Correspondingly, the payoff of a cooperator and a defector can be directly written as
$P_{C}(t) = \frac{b_{\rm m}y}{R_{\rm m}}-g$ and $P_{D}(t) = \frac{b_{\rm m}y}{R_{\rm m}} - p\beta$, respectively. Accordingly, we can obtain the equation system with time delay as
\begin{equation}
\left\{
\begin{array}{ll}
\dot{x}=x(1-x)(p\beta-g)\\
\dot{y}=ry(1-\frac{y}{R_{\rm m}})-N\frac{y}{R_{\rm m}}b_{\rm m}+gNx(t-\tau).\\
\end{array}
\right.
\label{85} \end{equation}

In the following, we study the possible equilibrium points and their stabilities of the above coupled equation system.
This equation system has at most three  meaningful fixed points when $p\beta\neq g$ which are
$(0,0)$, $(0,R_{\rm m}-\frac{Nb_{\rm m}}{r})$,  and $(1,\frac{R_{\rm m}r-Nb_{\rm m}+\sqrt{G}}{2r})$, where $G=(R_{\rm m}r-Nb_{\rm m})^{2}+4rR_{\rm m}gN$. We respective use $F_0$, $F_1$, and $F_2$ to represent these three fixed points. In the special case of $p\beta=g$, we find that the fixed point in the system depending on the initial conditions is $(x_{0},\frac{R_{\rm m}r-Nb_{\rm m}+\sqrt{Gx_{0}}}{2r})$ denoted by $F_3$
where the initial conditions and history functions for equation (\ref{85}) are assumed to be
\begin{equation}
\begin{array}{ll}
x(\xi)=x_{0}, y(\xi)=y_{0} \\
x_{0}\geq0, y_{0}\geq0, \xi\in[-\tau,0],\\
\end{array}
\label{86} \end{equation}
where $(x_{0}, y_{0})\in [0,1]\times[0,R_{\rm m}]$.

Next we use the method of characteristic roots of delay differential equations to study stabilities of these fixed points and obtain the following theorem.

\newtheorem{mythm}{Theorem}
\begin{mythm}\label{thm:point}
 Suppose that the fixed point of equation (\ref{85}) is $(x^{*}, y^{*})$.

(1) The stability of the fixed point $(x^{*}, y^{*})$ is irrelevant with time delay.

(2) The characteristic roots of equation (\ref{85}) are $\lambda_{1}=r-\frac{2ry^{*}}{R_{\rm m}}-\frac{Nb_{\rm m}}{R_{\rm m}}$
and $\lambda_{2}=(1-2x^{*})(p\beta-g)$, respectively.
\end{mythm}

{\bf Proof.}
(1) The linearized equation of equation (\ref{85}) at a fixed point defined by $(x^{*},y^{*})$ is
\begin{equation}
\left\{
\begin{array}{ll}
\dot{x}=x(1-2x^{*})(p\beta-g)\\
\dot{y}=y(r-\frac{2ry^{*}}{R_{\rm m}}-\frac{Nb_{\rm m}}{R_{\rm m}})+gNx(t-\tau).\\
\end{array}
\right.\label{87}
\end{equation}
Accordingly, the characteristic equation of equation (\ref{87}) can be written as
\begin{equation}
\lambda^{2}+a\lambda+b\lambda {\rm e^{-\lambda\tau}}+c+d{\rm e^{-\lambda\tau}}=0,
\end{equation}
where $a=-[r-\frac{2ry^{*}}{R_{\rm m}}-\frac{Nb_{\rm m}}{R_{\rm m}}+(1-2x^{*})(p\beta-g)]$, $b=0$,
$c=(1-2x^{*})(p\beta-g)(r-\frac{2ry^{*}}{R_{\rm m}}-\frac{Nb_{\rm m}}{R_{\rm m}})$, and $d=0$.
Since $b^{2}+2c-a^{2}<0$ and $c^{2}-d^{2}>0$, there does not exist the purely imaginary solution and there are no stability switches for any $\tau\geq0$ \cite{kuang_93}. Therefore, the stability of the fixed point $(x^{*}, y^{*})$
for $\tau>0$
is the same with $\tau=0$, and the stability of the fixed point $(x^{*}, y^{*})$ is irrelevant with time delay.
 Accordingly, by means of the Hartman-Grobman
Theorem~\cite{perko_01}, the stability of these fixed points in equation~(\ref{85})
is irrelevant with time delay, which indicates that the evolutionary outcome of the system is irrelevant with time delay.

(2) Since the characteristic equation is a usual quadratic equation, it has two roots at most, independent of time delay, which are $\lambda_{1}=r-\frac{2ry^{*}}{R_{\rm m}}-\frac{Nb_{\rm m}}{R_{\rm m}}$
and $\lambda_{2}=(1-2x^{*})(p\beta-g)$, respectively.

In the following, we distinguish two substantially different parameter
regions where the distinction is based on
the actual intrinsic growth rate value of the renewable common pool resource.\\

\subsection*{\rm{\textbf{A1. Slowly growing resource pool}}}

First,  we consider the case in which the resource
pool is recovering slowly due to small intrinsic growth rate, which assumes that $0<r\leq\frac{Nb_{\rm m}}{R_{\rm m}}$. In this situation, the system has only two fixed points in the parameter space of $0\leq x\leq1$ and $y \geq 0$.
They are $F_{0}$ and $F_{2}$, respectively. In dependence of the efficiency of inspection and punishment, we can distinguish two main cases. Note that the combined effect of these institutions can be characterized by the product of $p$ and $\beta$ parameters. The first case is when they are efficient, hence $p\beta$ exceeds $g$. According to Theorem~1, we know that the stability of these two fixed points is  irrelevant with time delay. Therefore, the stability of these two fixed points for $\tau>0$ is the same with $\tau=0$. Moreover, these eigenvalues of the fixed point $F_{2}$ for $\tau=0$ are negative, whereas the largest eigenvalue of the fixed point $F_{0}$ is positive.
Consequently, the fixed point $F_{0}$ is unstable, while the fixed point $F_{2}$ is asymptotically stable. Therefore, the fixed point $F_{2}$ is asymptotically stable for $\tau\geq0$.

\begin{figure*}
\centering
\fbox{ \begin{minipage}{6.0 in}
\centering
\includegraphics[width=4in]{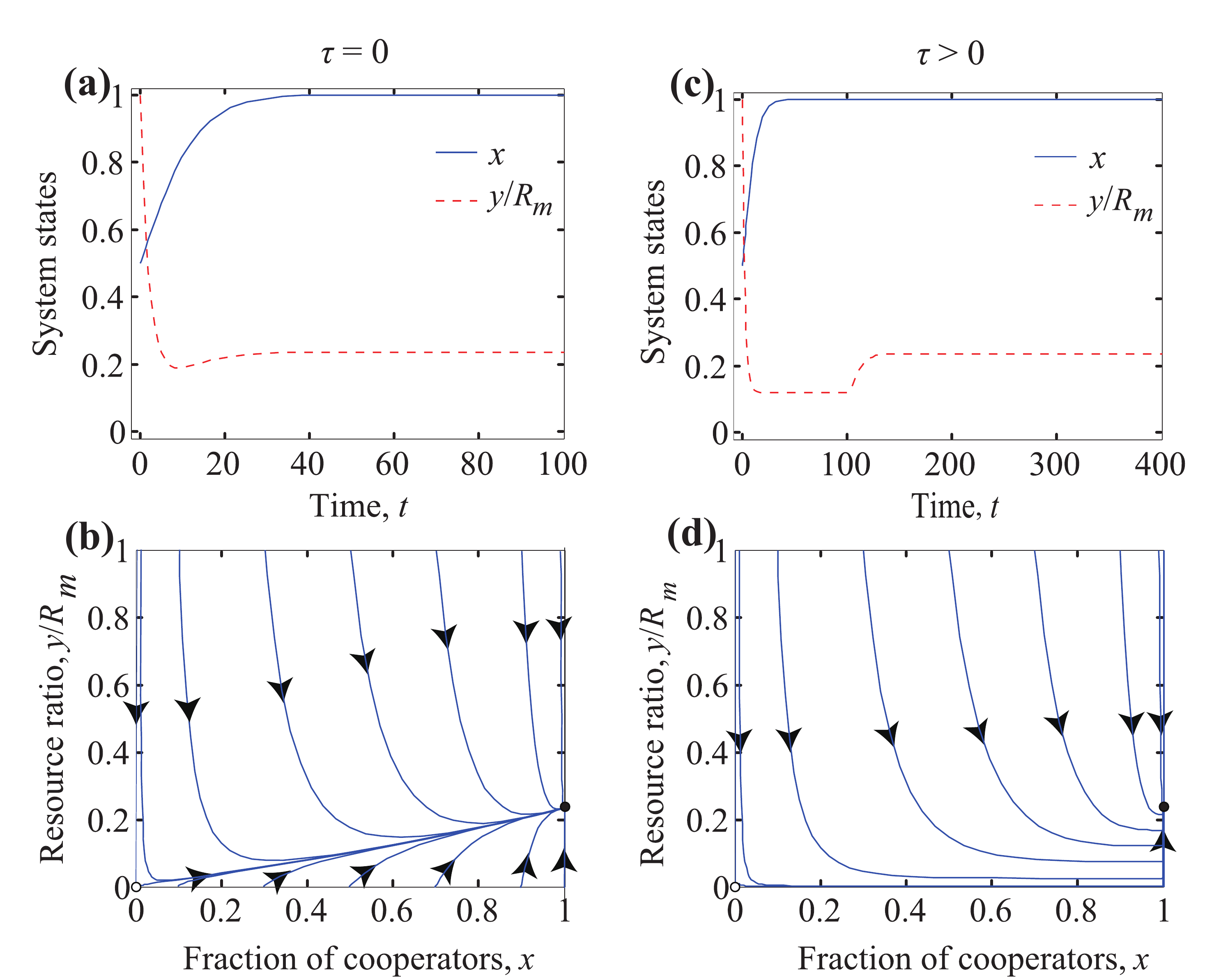}
\caption{Coevolutionary dynamics  for  $r <\frac{Nb_{\rm m}}{R_{\rm m}} $ and $p\beta>g$. Top panels show the time evolution of the fraction of cooperators and the resource ratio. Bottom panels show the related phase  portrait on $x - y/Rm $ plane. Filled (open) circle represents a stable (unstable) fixed point.  Parameters are $r = 0.1$, $N = 1000$, $g=0.1$, $R_{\rm m} = 1000$, $p = 0.5$, $\beta = 0.5$, and $b_{\rm m} = 0.5$. The applied time delay in the left column is $\tau=0$, while in the right column is $\tau=100$.}\label{figs1}
\end{minipage}}
\end{figure*}

The coevolutionary dynamics for  $0<r<\frac{Nb_{\rm m}}{R_{\rm m}}$ and $p\beta>g$ are plotted in figure \ref{figs1}. We can see that when the product of $p\beta$ exceeds $g$, no matter whether $\tau$ is 0 or 100, the system will eventually reach the same $F_{2}$ state, which means that the evolutionary outcome of the system is irrelevant with time delay. As time increases, cooperators become more and more until they dominate the population. At the same time, the resource value is positive, which ensures sustainability.

The second case is when the inspection-punishment institutions are less effective and the term $g$
exceeds $p\beta$ products. In this case, the system described by equation (\ref{85}) also has the same two fixed points, which are $F_{0}$ and $F_{2}$. According to Theorem~1, we know that the stability of these two fixed points is irrelevant with time delay. Therefore, the stability of these two fixed points for $\tau>0$ is the same with $\tau=0$. Moreover, these eigenvalues of the fixed point $F_{2}$ for $\tau=0$ are positive, whereas the largest eigenvalue of the fixed point $F_{0}$ is negative. Consequently, $F_{2}$ is unstable, while $F_{0}$ is asymptotically stable for all $\tau\geq0$.

\begin{figure*}
\centering
\fbox{ \begin{minipage}{6.0 in}
\centering
\includegraphics[width=4in]{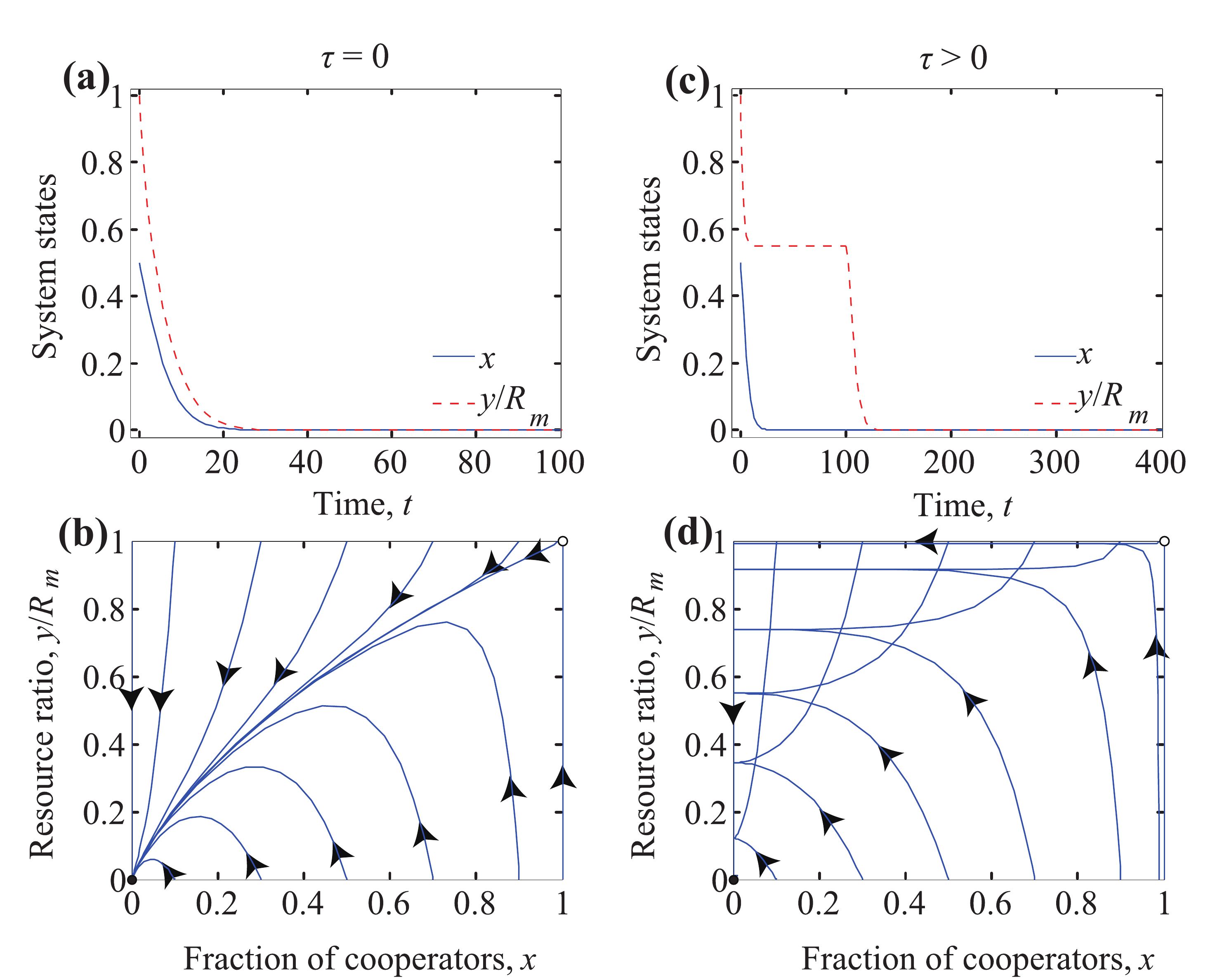}
\caption{Coevolutionary dynamics  for  $r <\frac{Nb_{\rm m}}{R_{\rm m}} $ and $p\beta<g$. Notations and the applied time delay values are the same as for figure~\ref{figs1}. Parameters are $r = 0.1$, $N = 1000$, $g=0.5$, $R_{\rm m} = 1000$, $p = 0.5$, $\beta = 0.5$, and $b_{\rm m} = 0.5$.}\label{figs2}
\end{minipage}}
\end{figure*}

The coevolutionary dynamics for $0<r < \frac{Nb_{\rm m}}{R_{\rm m}} $ and $0<p\beta<g$ are plotted in figure \ref{figs2}. We can see that no matter whether $\tau$ is 0 or 100, the system converges to the fixed point $F_{0}$, which means that the evolutionary outcome of the system is irrelevant with time delay. This suggests that resources become less and less until they are exhausted, but first cooperators become extinct. That is to say, when the inspection-punishment institutions are less effective, defectors dominate leading to the depletion of common resources.

In the special case of $p\beta=g$, we have $\dot{x}=0$. Accordingly, the equation system becomes
\begin{equation}
\left\{
\begin{array}{ll}
\dot{x}=0\\
\dot{y}=ry(1-\frac{y}{R_{\rm m}})-N\frac{y}{R_{\rm m}}b_{\rm m}+gNx(t-\tau).\\
\end{array}
\right.
\label{88} \end{equation}
Here the fixed point in the system depends on its initial conditions, which is $F_{3}(x_{0},\frac{R_{\rm m}r-Nb_{\rm m}+\sqrt{Gx_{0}}}{2r})$.
The corresponding characteristic equation for the eigenvalues $\lambda$ is
\begin{equation}
\lambda^{2}-(r-\frac{2ry^{*}}{R_{\rm m}}-\frac{Nb_{\rm m}}{R_{\rm m}})\lambda=0.
\label{89} \end{equation}
According to Theorem~1, the stability of fixed points is irrelevant with time delay, therefore the stability of related fixed points for $\tau>0$ is the same with $\tau=0$. Since $\lambda(\tau)=0$ is always a root of equation (\ref{89}) for $\tau=0$, the fixed point $F_{3}$
is stable, but not asymptotically stable \cite{kuang_93}.

Furthermore, we provide the theoretical analysis of the equilibrium points for the special case of $r=\frac{Nb_{\rm m}}{R_{\rm m}}$. In dependence of the efficiency of inspection and punishment, we can further distinguish three following sub-cases.

In the first case of $p\beta>g$, the equation system becomes
\begin{equation}
\left\{
\begin{array}{ll}
\dot{x}=x(1-x)(p\beta-g)\\
\dot{y}=-\frac{Nb_{\rm m}y^{2}}{R_{\rm m}^{2}}+gNx(t-\tau).\\
\end{array}
\right.
\label{90} \end{equation}

The corresponding characteristic equation for the eigenvalues $\lambda$ at the fixed point $(x^{*},y^{*})$ is
\begin{equation}
\lambda^{2}+[\frac{2Nb_{\rm m}y^{*}}{R^2_{\rm m}}+(1-2x^{*})(p\beta-g)]\lambda+\frac{2Nb_{\rm m}y^{*}}{R^2_{\rm m}}(1-2x^{*})(p\beta-g)=0.
\label{91} \end{equation}

Then the equation system has two fixed points which are $F_{0}$ and $F_{2}$, respectively. According to Theorem~1, we know that the stability of these two fixed points is  irrelevant with time delay. Therefore, the stability of these two fixed points for $\tau>0$ is the same with $\tau=0$. Moreover, these eigenvalues of the fixed point $F_{2}$ for $\tau=0$ are negative, whereas the largest eigenvalue of the fixed point $F_{0}$ is positive. Consequently, the fixed point $F_{0}$ is unstable, while the fixed point $F_{2}$ is asymptotically stable for $\tau\geq0$.

In the second case of $p\beta<g$, the equation system can also be depicted by equation (\ref{90}) and accordingly has two fixed points, which are $F_{0}$ and $F_{2}$, respectively. According to Theorem~1, the stability of these two fixed points for $\tau>0$ is the same with $\tau=0$. Here the largest eigenvalue of the fixed point $F_{2}$ is positive, therefore it is unstable. Regarding $F_{0}$,
since $\lambda(\tau)=0$ is always a root of equation (\ref{91}), the fixed point $F_{0}$ is stable, but not asymptotically stable \cite{kuang_93}.

In the third case of $p\beta=g$, we have $\dot{x}=0$ and the equation system becomes
\begin{equation}
\left\{
\begin{array}{ll}
\dot{x}=0\\
\dot{y}=-\frac{Nb_{\rm m}y^{2}}{R_{\rm m}^{2}}+gNx(t-\tau).\\
\end{array}
\right.
\label{92} \end{equation}
Here the fixed point is $F_{3}$, which depends on its initial conditions. The corresponding characteristic equation for the eigenvalues $\lambda$ is
\begin{equation}
\lambda^{2}+\frac{2Nb_{\rm m}y^{*}}{R^2_{m}}\lambda=0.
\label{93} \end{equation}
Based on Theorem~1 we need to study the stability of fixed point at $\tau=0$. Since $\lambda(\tau)=0$ is always a root of equation (\ref{93}) for  $\tau=0$,
$F_{3}$ is stable, but not asymptotically stable for $\tau\geq0$~\cite{kuang_93}.\\

\subsection*{\rm{\textbf{A2. Rapidly growing resource pool}}}

If the intrinsic growth rate of resource pool becomes faster, means $r>\frac{Nb_{\rm m}}{R_{\rm m}}$, then the conclusion is more subtle. In this situation, the system described by equation (\ref{85}) has three fixed points, which are $F_{0}$, $F_{1}$, and $F_{2}$, respectively. As previously, we study the stability of fixed point at $\tau=0$. According to the sign of the largest eigenvalue, $F_{0}$ is unstable for $\tau\geq0$. For the remaining two fixed points $F_{1}$ and $F_{2}$, their stabilities depend on the efficiency of inspection and punishment. When they are effective and the product of $p\beta$ exceeds $g$, the fixed point $F_{2}$ is asymptotically stable for $\tau\geq0$, while $F_{1}$ is unstable.

\begin{figure*}
\centering
\fbox{ \begin{minipage}{6.0 in}
\centering
\includegraphics[width=4.0in]{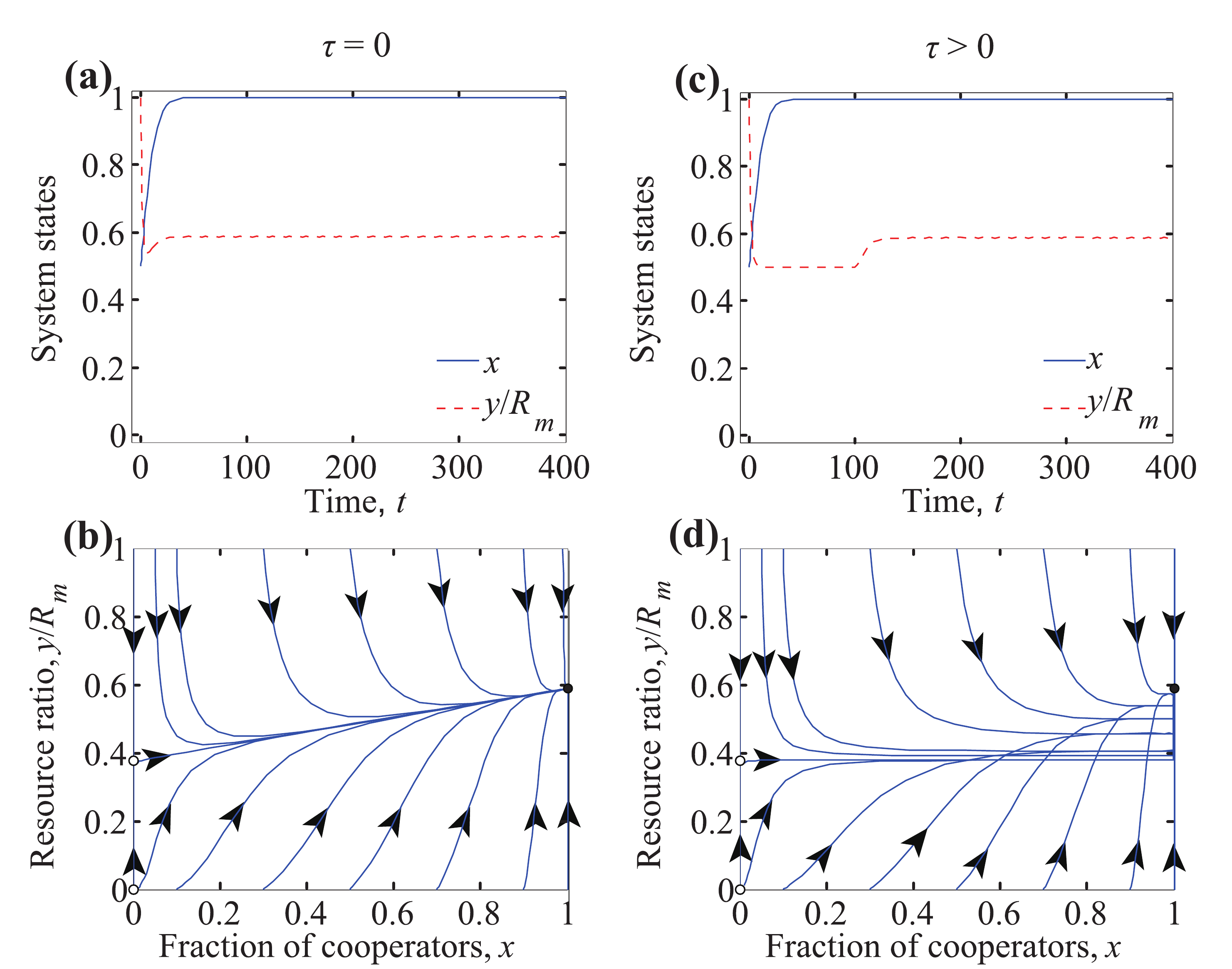}
\caption{Coevolutionary dynamics  for  $r >\frac{Nb_{\rm m}}{R_{\rm m}} $ and $p\beta>g$. Notations and the applied time delay values are the same as earlier. Parameters are $r = 0.8$, $N = 1000$, $g=0.1$, $R_{\rm m} = 1000$, $p = 0.5$, $\beta = 0.5$, and $b_{\rm m} = 0.5$.}\label{figs3}
\end{minipage}}
\end{figure*}

The coevolutionary dynamics for $r >\frac{Nb_{\rm m}}{R_{\rm m}} $ and $p\beta>g$ are plotted in figure~\ref{figs3}. It demonstrates that independently of the value of $\tau$, the system converges to the $F_{2}$ fixed point, signaling that time delayed feedback has no impact on the evolutionary outcome and the system terminates onto a sustainable state.

\begin{figure*}
\centering
\fbox{ \begin{minipage}{6.1 in}
\centering
\includegraphics[width=5.0in]{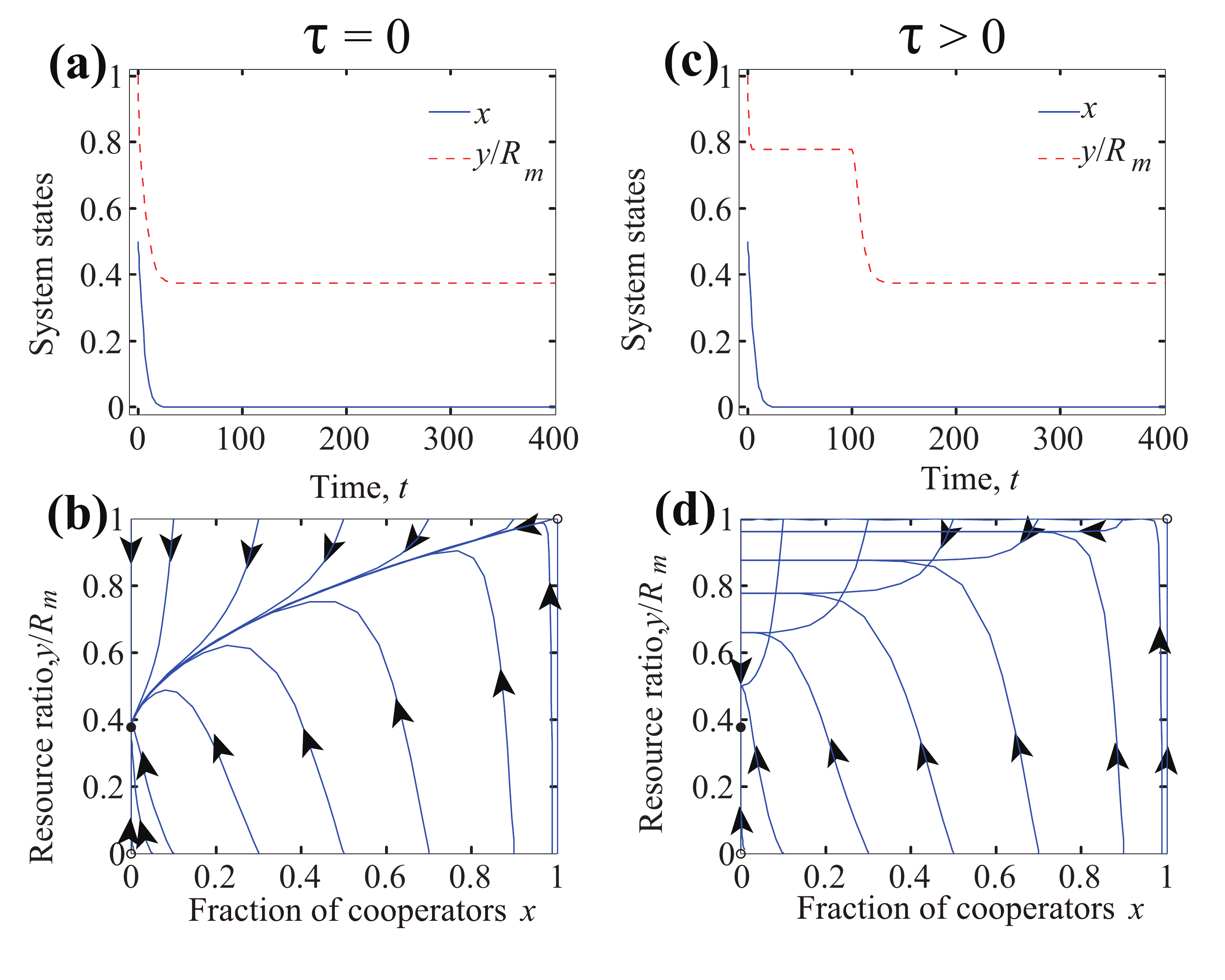}
\caption{Coevolutionary dynamics  for $r >\frac{Nb_{\rm m}}{R_{\rm m}} $ and $p\beta<g$. Notations and applied $\tau$ values are the same as previously. Parameters are $r = 0.8$, $N = 1000$, $g=0.5$, $R_{\rm m} = 1000$, $p = 0.5$, $\beta = 0.5$, and $b_{\rm m} = 0.5$.}\label{figs4}
\end{minipage}}
\end{figure*}

If the above mentioned institutions are less powerful, then the product $p\beta$ is less than the $g$ value. The results are opposite, which means that the fixed point $F_{1}$ is asymptotically stable for $\tau\geq0$, while $F_{2}$ is unstable. The trajectories of related dynamics are plotted in  figure~\ref{figs4}. We can see that no matter whether $\tau$ is 0 or 100, the system converges to the fixed point $F_{1}$, which means that the evolutionary outcome is irrelevant of time delay. This suggests that the system will reach a full defection state. Still, the latter is a sustainable state because the strong growing capacity of resource is capable to compensate to greediness of defective players.

In the special case when $p\beta=g$, we have $\dot{x}=0$, yielding a fixed point $F_{3}$.
The corresponding characteristic equation for the eigenvalue $\lambda$ is
\begin{equation}
\lambda^{2}-(r-\frac{2ry^{*}}{R_{\rm m}}-\frac{Nb_{\rm m}}{R_{\rm m}})\lambda=0.
\label{94} \end{equation}
According to the sign of the largest eigenvalue, $F_{0}$ is unstable for $\tau\geq0$. Furthermore, since $\lambda(\tau)=0$ is always a root of equation (\ref{94}) for  $\tau=0$, $F_{3}$ is stable, but not asymptotically stable for $\tau\geq0$~\cite{kuang_93}.\\

\section*{Appendix B. Feedback-evolving game with time delay by using proportional contribution}
\setcounter{equation}{0}
\renewcommand\theequation{B.\arabic{equation}}
In the remaining main section, we assume that cooperators' contribution to the common pool is proportional to their endowment. Accordingly, the dynamical equations of the coupled resource-strategy system are
\begin{equation}
\left\{
\begin{array}{ll}
\dot{x}=x(1-x)(p\beta-\frac{\alpha b_{\rm m}y}{R_{\rm m}})\\
\dot{y}=ry(1-\frac{y}{R_{\rm m}})-N\frac{y}{R_{\rm m}}b_{\rm m}+\frac{\alpha Nb_{\rm m}x(t-\tau)y(t-\tau)}{R_{\rm m}}\,\,.\\
\end{array}
\right.
\label{1}
\end{equation}
This equation system has at most five fixed points, which are $(0,0)$, $(1,0)$, $(0,R_{\rm m}-\frac{Nb_{\rm m}}{r})$,
$(1,R_{\rm m}-\frac{Nb_{\rm m}(1-\alpha)}{r})$,  and
$(K,\frac{R_{\rm m}p\beta}{\alpha b_{\rm m}})$, respectively,
where $K=\frac{1}{\alpha}-\frac{rR_{\rm m}}{\alpha b_{\rm m}N}+\frac{p\beta R_{\rm m}r}{N\alpha^{2}b_{\rm m}^{2}}$. We use $F_0$, $F_1$, $F_2$, $F_3$, and $F_4$ to represent these five fixed points, respectively.

The linearized equation of equation (\ref{1}) at a fixed point defined by $(x^{*},y^{*})$ is
\begin{equation}
\left\{
\begin{array}{ll}
\dot{x}=x(1-2x^{*})(p\beta-\frac{\alpha b_{\rm m}y^{*}}{R_{\rm m}})+x^{*}(x^{*}-1)\frac{\alpha b_{\rm m}y}{R_{\rm m}}\\
\dot{y}=y(r-\frac{2ry^{*}}{R_{\rm m}}-\frac{Nb_{\rm m}}{R_{\rm m}})+\frac{\alpha Nb_{\rm m}y^{*}x(t-\tau)}{R_{\rm m}}+\frac{\alpha Nb_{\rm m}x^{*}y(t-\tau)}{R_{\rm m}}.\\
\end{array}
\right.
\label{2}
\end{equation}
Accordingly, the characteristic equation of equation (\ref{2}) can be written as
\begin{equation}
\lambda^{2}+a\lambda+b\lambda \rm e^{-\lambda \tau}+c+d\rm e^{-\lambda\tau}=0,
\label{4}
\end{equation}
where $a=-[r-\frac{2ry^{*}}{R_{\rm m}}-\frac{Nb_{\rm m}}{R_{\rm m}}+(1-2x^{*})(p\beta-\frac{\alpha b_{\rm m}y^{*}}{R_{\rm m}})]$,
$b=-\frac{N\alpha b_{\rm m}x^{*}}{R_{\rm m}}$,
$c=(1-2x^{*})(p\beta-\frac{\alpha b_{\rm m}y^{*}}{R_{\rm m}})(r-\frac{2ry^{*}}{R_{\rm m}}-\frac{Nb_{\rm m}}{R_{\rm m}})$,
and
$d=(1-2x^{*})(p\beta-\frac{\alpha b_{\rm m}y^{*}}{R_{\rm m}})\frac{N\alpha b_{\rm m}x^{*}}{R_{\rm m}}-\frac{N\alpha^{2}b_{\rm m}^{2}}{R_{\rm m}^{2}}y^{*}x^{*}(x^{*}-1)$. Since it contains the term $\rm e^{-\lambda \tau}$  for $d\neq0$, it is a transcendental equation which has infinite roots. Note that stability changes of the fixed point $(x^{*},y^{*})$ can only occur for $\lambda=\rm i\omega$.
By substituting $\lambda=\rm i\omega$ into equation (\ref{4}) and by extracting the real and imaginary parts, we get the following equations
\begin{equation}
c-\omega^{2}+b\omega \sin\omega\tau+d\cos\omega\tau=0
\label{5} \end{equation}
and
\begin{equation}
a\omega+b\omega \cos\omega\tau-d\sin\omega\tau=0.
\label{6} \end{equation}
Thus, we have
\begin{equation}
\omega^{4}-H\omega^{2}+c^{2}-d^{2}=0,
\label{7} \end{equation}
where $H=b^{2}+2c-a^{2}.$ Its roots are
\begin{equation}
\omega_{\pm}^{2}=\frac{1}{2}[H\pm (H^{2}-4(c^{2}-d^{2}))^{\frac{1}{2}}]\,.
\label{8} \end{equation}
We can then obtain the existence condition of the imaginary root $\lambda=\rm i\omega$ with $\omega>0$, which can be written in the following proposition.

\newtheorem{mypro}{Proposition}
\begin{mypro}\label{thm:point}
There is only one imaginary solution
$\lambda=\rm i\omega_{+}$ with $\omega_{+}>0$, if one of the following three conditions holds: (\textrm{1}) $c^{2}<d^{2}$;
(\textrm{2}) $c^{2}=d^{2}$ and $H>0$;
(\textrm{3}) $H^{2}-4(c^{2}-d^{2})=0$ and $H>0$.
There are two imaginary solutions,
$\lambda_{\pm}=\rm i\omega_{\pm}$, with $\omega_{+}>\omega_{-}>0$, if $c^{2}>d^{2}$, $H>0$, and $H^{2}-4(c^{2}-d^{2})>0$. Otherwise, there are no imaginary solutions.
\end{mypro}

{\bf Proof.} In equation (\ref{8}), if $c^{2}<d^{2}$, then $H^{2}-4(c^{2}-d^{2})>H^{2}$.
Accordingly, $H+ (H^{2}-4(c^{2}-d^{2}))^{\frac{1}{2}}>0$ and $H- (H^{2}-4(c^{2}-d^{2}))^{\frac{1}{2}}<0$,
therefore there is only one imaginary solution $\lambda=\rm i\omega_{+}$
with $\omega_{+}={\frac{1}{2}[{{H+ (H^{2}-4(c^{2}-d^{2}))^{\frac{1}{2}}}}]}^{\frac{1}{2}}$.

If $c^{2}=d^{2}$ and $H>0$, then  $H+ (H^{2}-4(c^{2}-d^{2}))^{\frac{1}{2}}=2H$
and $H- (H^{2}-4(c^{2}-d^{2}))^{\frac{1}{2}}=0$.
Based on the condition $\omega>0$, $H- (H^{2}-4(c^{2}-d^{2}))^{\frac{1}{2}}=0$ is not satisfied.
Accordingly, $\omega_{+}=\sqrt{2H}$, and there is only one imaginary solution, namely $\lambda=\rm i\omega_{+}$ with $\omega_{+}=\sqrt{2H}$.

If $H^{2}-4(c^{2}-d^{2})=0$ and $H>0$, then
$H+ (H^{2}-4(c^{2}-d^{2}))^{\frac{1}{2}}=H- (H^{2}-4(c^{2}-d^{2}))^{\frac{1}{2}}=H$.
Accordingly, $\omega_{+}=\sqrt{H}$. Hence there is only one
imaginary solution, which is $\lambda=\rm i\omega_{+}$ with $\omega_{+}=\sqrt{H}$.

If $c^{2}>d^{2}$, $H>0$, and $H^{2}-4(c^{2}-d^{2})>0$,
then $H^{2}-4(c^{2}-d^{2})<H^{2}$. Accordingly, $H+ (H^{2}-4(c^{2}-d^{2}))^{\frac{1}{2}}>0$ and $H- (H^{2}-4(c^{2}-d^{2}))^{\frac{1}{2}}>0$. Therefore, there are two imaginary solutions $\lambda_{\pm}=\rm i\omega_{\pm}$ with
$\omega_{\pm}={\frac{1}{2}[{{H\pm (H^{2}-4(c^{2}-d^{2}))^{\frac{1}{2}}}}]}^{\frac{1}{2}}$.

The proof for the case of no imaginary solutions is similar, which can be found in reference~\cite{cao_j_ieee07}. For further analysis of our equation system with time delay, let us denote $e_{C}=\frac{Nb_{\rm m}(1-\alpha)}{R_{\rm m}}$ and $e_{D}=\frac{Nb_{\rm m}}{R_{\rm m}}$, respectively, representing the gain rates of cooperators and defectors in a population from the common resource \cite{chen_xj_pcb18}.
It also involves that we have $0\leq e_{C}\leq e_{D} \leq 1$. In the following, we distinguish three significantly different parameter regions where the distinction is based on the actual intrinsic growth rate value of the renewable common pool resource.\\

\subsection*{\rm{\textbf{B1. Slowly growing resource pool}}}

In the case of $0<r<e_{C}<e_{D}$, the equation system has just two fixed points which are $F_{0}$ and $F_{1}$, respectively.
The characteristic equation of equation (\ref{2}) at the fixed point $F_{0}$ can be written as

\begin{equation}
\lambda^{2}-(r-e_{D}+p\beta)\lambda+p\beta(r-e_{D})=0\,.
\label{9} \end{equation}
Since $H<0$ and $c^{2}-d^{2}>0$, there is no purely imaginary solution and the stability of the fixed point $F_{0}$ for $\tau>0$ is the same with $\tau=0$ according to Proposition 1. Moreover, these eigenvalues of the fixed point $F_{0}$ for $\tau=0$
are $\lambda_{1}=r-e_{D}<0$ and $\lambda_{2}=p\beta>0$, respectively, which means that $F_{0}$ is unstable for $\tau=0$. Therefore, $F_{0}$ is unstable for $\tau\geq0$ \cite{kuang_93}.

The characteristic equation of equation (\ref{2}) at the fixed point $F_{1}$ is
\begin{equation}
\lambda^{2}-(r-e_{D}-p\beta)\lambda-\alpha e_{D} \lambda {\rm e}^{-\lambda\tau}-p\beta(r-e_{D})-p\beta\alpha e_{D}\rm e^{-\lambda \tau} =0\,.
\label{10} \end{equation}
Since $H<0$, $c^{2}-d^{2}>0$, and these eigenvalues of the fixed point $F_{1}$ for $\tau=0$ are $\lambda_{1}=-p\beta<0$ and $\lambda_{2}=r-e_{C}<0$, respectively, $F_{1}$ is asymptotically stable for $\tau>0$ according to Proposition 1 and reference~\cite{kuang_93}.

For the special case of $r=e_{C}$, we have $r=\frac{Nb_{\rm m}(1-\alpha)}{R_{\rm m}}$.
Then the equation system has two fixed points in the parameter space, which are $F_{0}$ and $F_{1}$, respectively. The characteristic equation of equation (\ref{2}) at the fixed point $F_{0}$ can be written as
\begin{equation}
\lambda^{2}-(r-e_{D}+p\beta)\lambda+p\beta(r-e_{D})=0.
\label{11} \end{equation}
Since $H<0$, $c^{2}-d^{2}>0$, and the eigenvalues of the fixed point $F_{0}$ for $\tau=0$ are $\lambda_{1}=r-e_{D}<0$ and $\lambda_{2}=p\beta>0$, $F_{0}$ is unstable for $\tau\geq0$ according to Proposition 1 and reference~\cite{kuang_93}.

The characteristic equation of equation (\ref{2}) at the fixed point $F_{1}$ is
\begin{equation}
\lambda^{2}-(r-e_{D}-p\beta)\lambda-\alpha e_{D} \lambda {\rm e}^{-\lambda\tau}-p\beta(r-e_{D})-p\beta\alpha e_{D}\rm e^{-\lambda \tau} =0.
\label{12} \end{equation}
In this situation, $a=p\beta-r+e_{D}>0$, $b=-\alpha e_{D}$, $c=-p\beta(r-e_{D})>0$,  and $d=-p\beta\alpha e_{D}\rm e^{-\lambda\tau}$.
Since $c+d=0$, $c=-p\beta(r-e_{D})>0$,  and $a^{2}-b^{2}=(p\beta-r+e_{C})(p\beta+e_{D}(1+\alpha)-r)>0$,
we obtain that $F_{1}$ is stable, but not asymptotically stable~\cite{kuang_93}.

\subsection*{\rm{\textbf{B2. Moderately growing resource pool}}}

In the case of $e_{C}<r<e_{D}$ and $\alpha b_{\rm m}(1-\frac{e_{C}}{r})> p\beta$,
the equation system has four fixed points, which are $F_{0}$, $F_{1}$, $F_{3}$, and $F_{4}$, respectively. The characteristic equation of equation (\ref{2}) at the fixed point $F_{0}$ is
\begin{equation}
\lambda^{2}-(r-e_{D}+p\beta)\lambda+p\beta(r-e_{D})=0\,.
\label{13} \end{equation}
Since $H<0$, $c^{2}-d^{2}>0$, and the eigenvalues of the fixed point for $\tau=0$ are $\lambda_{1}=r-e_{D}<0$ and $\lambda_{2}=p\beta>0$, $F_{0}$ is unstable for $\tau\geq0$ according to Proposition 1 and reference~\cite{kuang_93}.

The characteristic equation of equation (\ref{2}) at the fixed point $F_{1}$ is
\begin{eqnarray}
\lambda^{2}-(r-e_{D}-p\beta)\lambda-\alpha e_{D} \lambda {\rm e}^{-\lambda\tau}-p\beta(r-e_{D})-p\beta\alpha e_{D}\rm e^{-\lambda \tau} =0.
\label{14} \end{eqnarray}
The corresponding eigenvalues are $\lambda_{1}=-p\beta<0$ and $\lambda_{2}=r-e_{C}>0$, yielding that the fixed point $F_{1}$ for $\tau=0$ is unstable.
We can further obtain
${\frac{d(R_{e}\lambda)}{d\tau}}|_{\lambda=\rm i\omega_{+}}>0$.
Therefore, according to reference~\cite{kuang_93}, $F_{1}$ remains unstable for $\tau\geq0$.

The characteristic equation of equation (\ref{2}) at the fixed point $F_{3}$ is
\begin{eqnarray}
 \lambda^{2}-(L+M)\lambda- \alpha e_{D}\lambda {\rm e}^{-\lambda\tau}
+ML+M\alpha e_{D}\rm e^{-\lambda \tau} =0,
\label{15} \end{eqnarray}
where $L=e_{D}-r-2\alpha e_{D}$ and $M=\alpha b_{\rm m}(1-\frac{e_{C}}{r})-p\beta$.
Since $H<0$ and $c^{2}-d^{2}>0$, and the eigenvalues for $\tau=0$ are $\lambda_{1}=\alpha b_{\rm m}(1-\frac{e_{C}}{r})-p\beta>0$ and $\lambda_{2}=e_{C}-r<0$, according to Proposition 1 and reference~\cite{kuang_93}, $F_{3}$ is unstable for  $\tau\geq0$.

For the fixed point $F_{4}$ the characteristic equation is
\begin{eqnarray}
\fl\lambda^{2}+(\frac{2rp\beta}{\alpha b_{\rm m}}+e_{D}-r)\lambda+(r-\frac{rp\beta}{\alpha b_{\rm m}}-e_{D})\lambda {\rm e}^{-\lambda\tau}
+Np\beta K(1-K)\frac{\alpha b_{\rm m}}{R_{\rm m}}\rm e^{-\lambda\tau}=0,
\label{16} \end{eqnarray}
where $K=\frac{1}{\alpha}-\frac{rR_{\rm m}}{\alpha b_{\rm m}N}+\frac{p\beta rR_{\rm m}}{N\alpha^{2}b_{\rm m}^{2}}$.
Since these eigenvalues for $\tau=0$ satisfy
$\lambda_{1}+\lambda_{2}<0$ and $\lambda_{1}\lambda_{2}>0$, $F_{4}$ is asymptotically stable for $\tau=0$.
Moreover, since $c^{2}<d^{2}$, according to Proposition 1 there is only one pair of purely imaginary solutions and $\lambda=\pm \rm i\omega_{+}$ with $\omega_{+}>0$,
where
\begin{eqnarray}
\omega_{+}^{2}=\frac{1}{2}\left[Q
+\left(Q^{2}+4Np\beta K
(1-K)\frac{\alpha b_{\rm m}}{R_{\rm m}}\right)^{\frac{1}{2}}\right]\,,
\label{17} \end{eqnarray}
with $Q=\frac{rp\beta}{\alpha b_{\rm m}}(2r-\frac{3rp\beta}{\alpha b_{\rm m}}-2e_{D})$.
According to reference~\cite{kuang_93}, there exists a critical time delay $\tau_{\rm c}=\frac{\theta_{1}}{\omega_{+}}$, where
$\theta_{1}$ satisfies
\begin{eqnarray}
\cos\theta_{1}=-\frac{(ab^{2}-d)\omega_{+}^{2}}{b^{2}\omega_{+}^{2}+d^{2}} \hspace{0.5cm}
\textrm{and} \hspace{0.5cm}
\sin\theta_{1}=-\frac{ad^{2}\omega_{+}+b\omega_{+}^{3}}{b^{2}\omega_{+}^{2}+d^{2}}.
\label{19} \end{eqnarray}
Hence $F_{4}$ is  asymptotically stable for $\tau<\tau_{\rm c}$ and unstable for $\tau>\tau_{\rm c}$.
For $\tau=\tau_{\rm c}$, there  exists a Hopf bifurcation and the direction and the stability of the bifurcating periodic solutions can be determined according to the following calculations.

We first do some transformation for the equation system as
$$x_{1}=x-x^{*},  x_{2}=y-y^{*},  t=\frac{t}{\tau}, \qquad{\rm and} \hspace{0.5cm} \tau=\tau_{\rm c}+\mu\,.$$
Accordingly, the equation system can be then written as the following form in $C=C([-1,0],R^{2})$.
\begin{eqnarray}
\dot{x}(t)=L_{\mu}(x_{t})+f(\mu, x_{t}),
\label{20} \end{eqnarray}
where $x(t)=(x_{1}(t), x_{2}(t))\in R^{2}$ and $L_{\mu}:C\rightarrow R^2$. And we have
\begin{eqnarray*}
L_{\mu}(\phi)
=(\tau_{\rm c}+\mu)\left[
\begin{array}{lll}
a_{11}&a_{12}\\
0&a_{22}\\
\end{array}\right]
\left[
\begin{array}{lll}
\phi_{1}(0)\\
\phi_{2}(0)\\
\end{array}\right]
+(\tau_{\rm c}+\mu)\left[
\begin{array}{lll}
0&0\\
b_{21}&b_{22}\\
\end{array}\right]
\left[
\begin{array}{lll}
\phi_{1}(-1)\\
\phi_{2}(-1)\\
\end{array}\right],
\end{eqnarray*}
where $a_{11}=(p\beta-\frac{\alpha b_{\rm m}y^{*}}{R_{\rm m}})(1-2x^{*})$, $a_{12}=\frac{x^{*}(x^{*}-1)\alpha b_{\rm m}}{R_{\rm m}}$,
$a_{22}=r-\frac{2ry^{*}}{R_{\rm m}}-\frac{Nb_{\rm m}}{R_{\rm m}}$, $b_{21}=\frac{N\alpha b_{\rm m}y^{*}}{R_{\rm m}}$,
$b_{22}=\frac{N\alpha b_{\rm m}x^{*}}{R_{\rm m}}$,
and $f:R\times C\rightarrow R^2$,
$f(\mu,\phi)=(\tau_{\rm c}+\mu)\left[
\begin{array}{lll}
f_{1}\\
f_{2}\\
\end{array}\right],$
with
$
f_{1}=\frac{-\alpha b_{\rm m}(1-2x^{*})}{R_{\rm m}}\phi_{1}(0)\phi_{2}(0)-(p\beta-\frac{\alpha b_{\rm m}y^{*}}{R_{\rm m}})(\phi_{1}(0))^{2}+\frac{\alpha b_{\rm m}}{R_{\rm m}}(\phi_{1}(0))^{2}\phi_{2}(0)
$
and
$f_{2}=\frac{-r}{R_{\rm m}}(\phi_{2}(0))^{2}+\frac{N\alpha b_{\rm m}}{R_{\rm m}}\phi_{1}(-1)\phi_{2}(-1)$,
where $\phi=(\phi_{1}, \phi_{2})\in C$.
By using Riesz representation theorem \cite{kuang_93}, there exists a function $\eta(\theta, \mu)$ of bounded variation for $\theta\in[-1,0]$, such that
$$L_{\mu}(\phi)=\int_{-1}^{0} d\eta(\theta,0)\phi(\theta)  \qquad{\rm for} \qquad \phi\in C.$$
In fact, we can take
\begin{eqnarray*}
L_{\mu}(\phi)
=(\tau_{\rm c}+\mu)\left[
\begin{array}{lll}
a_{11}&a_{12}\\
0&a_{22}\\
\end{array}\right]
\delta(\theta)
+(\tau_{\rm c}+\mu)\left[
\begin{array}{lll}
0&0\\
b_{21}&b_{22}\\
\end{array}\right]
\delta(\theta+1),
\end{eqnarray*}
where $\delta$ is the Dirac delta function defined as
$$\delta(\theta)=\left\{\begin{array}{ll}
0,&\theta\neq0,\\
1,&\theta=0.\\
\end{array}\right.$$

For $\phi\in C^{1}([-1,0],R^{2})$, we respectively define
$$A(\mu)\phi=\left\{\begin{array}{ll}
\qquad\frac{d\phi(\theta)}{d\theta},&\theta\in[-1, 0),\\
\int_{-1}^{0} d\eta(\mu, \theta)\phi(\theta),&\theta=0,\\
\end{array}\right.$$
and
$$R(\mu)\phi=\left\{\begin{array}{ll}
0,&\theta\in[-1, 0),\\
f(\mu, \phi),&\theta=0.\\
\end{array}\right.$$
Then the system described by equation (\ref{1}) is equivalent to
\begin{eqnarray}
\dot{x}_{t}=A(\mu)x_{t}+R(\mu)x_{t},
\label{21} \end{eqnarray}
where $x_{t}(\theta)=x(t+\theta)$ for $\theta\in[-1, 0)$.

For $\psi\in C^{1}([0,1],(R^{2})^{*})$, we respectively define
$$A^{*}\psi(s)=\left\{\begin{array}{ll}
\qquad -\frac{d\psi(s)}{ds},&s\in(0, 1],\\
\int_{-1}^{0} d\eta^{T}(t, 0)\psi(-t),&s=0,\\
\end{array}\right.$$
and a bilinear inner product
\begin{eqnarray}
<\psi(s), \phi(\theta)>=\bar{\psi}(0)\phi(0)-\int_{-1}^{0}\int_{\zeta=0}^{\theta} \bar{\psi}(\zeta-\theta)d \eta(\theta)\phi(\zeta)d \zeta,
\label{22} \end{eqnarray}
where $\eta(\theta)=\eta(\theta,0)$. Then $A(0)$ and $A^{*}$ are adjoint operators. Suppose that $q(\theta)$ and $q^{*}(s)$ are
eigenvectors of $A$ and $A^{*}$ corresponding to $\rm i\omega\tau_{\rm c}$ and $-\rm i\omega\tau_{\rm c}$, respectively.
Then  $q(\theta)=(1,q_{1})^{T}\rm exp(\rm i\omega\tau_{\rm c}\theta)$
is the eigenvector of $A(0)$ corresponding to $\rm i\omega\tau_{\rm c}$, and $A(0)q(\theta)=\rm i\omega\tau_{\rm c}q(\theta)$.
It follows from the definitions of $A(0)$, $L_{\mu}\phi$, and $\eta(\theta, \mu)$ that
$$\tau_{\rm c}\left[
\begin{array}{lll}
{\rm i}\omega_{0}-a_{11}&-a_{12}\\
-b_{21}{\rm e^{-{\rm i}\omega\tau_{\rm c}}}&{\rm i}\omega_{0}-a_{22}-b_{22}{\rm e^{-\rm i\omega\tau_{\rm c}}}\\
\end{array}\right]
\left[
\begin{array}{lll}
1\\
q_{1}\\
\end{array}\right]
=\left[
\begin{array}{lll}
0\\
0\\
\end{array}\right].
$$
Thus we can easily get
$q(\theta)=(1,q_{1})^{T}\rm e^{\rm i\omega\tau_{\rm c}\theta}$,
where $q_{1}=\frac{{\rm i}\omega-a_{11}}{a_{12}}$.

Similarly, let $q^{*}(s)=D(1,q_{2})\rm exp(\rm i\omega\tau_{\rm c}s)$ be the eigenvector of  $A^{*}$ corresponding to $-\rm i\omega\tau_{\rm c}$.
Based on $A^{*}$, we can obtain
 $q_{2}=-\frac{a_{12}}{{\rm i}\omega+a_{22}+b_{22}\rm exp(\rm i\omega\tau_{\rm c})}$.

To satisfy that $<q^{*}(s), q(\theta)>=1$, we need to evaluate the value of $D$. From the definition of the bilinear inner product
\begin{eqnarray*}
\fl<q^{*}(s), q(\theta)>
&=&\bar{D}(1,\bar{q}_{2})(1,q_{1})^{T}
-\int_{-1}^{0}\int_{\zeta=0}^{\theta} \bar{D}(1,\bar{q}_{2}){\rm exp(-\rm i\omega\tau_{\rm c}(\zeta-\theta))}d\eta(\theta)(1,q_{1})^{T}{\rm exp(\rm i\omega\tau_{\rm c}\zeta)} d\zeta\\
&=&\bar{D}[1+\bar{q}_{2}q_{1}-\int_{-1}^{0}(1,\bar{q}_{2})\theta {\rm exp(-\rm i\omega\tau_{\rm c})} d\eta(\theta)(1,q_{1})^{T}
]\\
&=&\bar{D}[1+\bar{q}_{2}q_{1}+\tau_{\rm c}(b_{21}\bar{q}_{2}+b_{22}q_{1}\bar{q}_{2}){\rm exp(-\rm i\omega\tau_{\rm c})}].
\end{eqnarray*}
We can thus choose $D$ as
$$D=\frac{1}{1+q_{2}\bar{q}_{1}+\tau_{\rm c}(b_{21}q_{2}+b_{22}\bar{q}_{1}q_{2})\rm exp(\rm i\omega\tau_{\rm c})},$$
such that  $<q^{*}(s), \bar{q}(\theta)>=0$.

In the following, we use the theorem by Hassard et al.~\cite{hassard_81} to compute the coordinates describing center manifold $C_{0}$ at $\mu=0$.
We then define
\begin{eqnarray}
z(t)=<q^{*}, x_{t}>  \qquad{\rm and}\qquad  W(t,\theta)=x_{t}(\theta)-2Re[z(t)q(\theta)].
\label{23} \end{eqnarray}
On the center manifold, we have
\begin{eqnarray*}
W(t,\theta)
&=& W(z(t), \bar{z}(t),\theta)\\
&=&W_{20}(\theta)\frac{z^{2}}{2}+W_{11}(\theta)z\bar{z}+W_{02}(\theta)\frac{\bar{z}^{2}}{2}+...,\\
\end{eqnarray*}
where $z$ and $\bar{z}$ are local coordinates for center manifold $C_{0}$ in the direction of $q$ and $\bar{q}^{*}$.
Note that $W$ is real if $x_{t}$ is real. We then only consider real solutions. For the solution $x_{t}\in C_{0}$,
since $\mu=0$, based on equation (\ref{20}) we have
\begin{eqnarray*}
\dot{z}
&=&{\rm i}\omega\tau_{\rm c}z+<q^{*}(\theta), f(0,  W(z, \bar{z},\theta)+2Re[zq(\theta)])>\\
&=&{\rm i}\omega\tau_{\rm c}z+\bar{q}^{*}(0)f(0,  W(z, \bar{z},0)+2Re[zq(0)])\\
&=&{\rm i}\omega\tau_{\rm c}z+\bar{q}^{*}(0)f_{0}(z, \bar{z})={\rm i}\omega\tau_{\rm c}z+g(z,\bar{z}),\\
\end{eqnarray*}
where
\begin{eqnarray}
\fl g(z,\bar{z})
= \bar{q}^{*}(0)f_{0}(z,\bar{z})\nonumber
=g_{20}(\theta)\frac{z^{2}}{2}+g_{11}(\theta)z\bar{z}+g_{02}(\theta)\frac{\bar{z}^{2}}{2}+g_{21}(\theta)\frac{\bar{z}^{2}\bar{z}}{2}+....\nonumber\\
\label{24}
\end{eqnarray}
By using equation (\ref{23}), we have
$x_{t}(x_{1t}(\theta), x_{2t}(\theta))=W(t,\theta)+zq(\theta)+\bar{z}q(\theta)$ and $q(\theta)=(1,q_{1})^{T}\rm exp(\rm i\omega\tau_{\rm c}\theta)$, and then
\begin{eqnarray*}
\fl x_{1t}(0)
=z+\bar{z}+W_{20}^{(1)}(0)\frac{z^{2}}{2}+W_{11}^{(1)}(0)z\bar{z}
+W_{02}^{(1)}(0)\frac{\bar{z}^{2}}{2}+O(|(z,\bar{z})|^{3}),\\
\fl x_{2t}(0)
=zq_{1}+\bar{z}\bar{q}_{1}+W_{20}^{(2)}(0)\frac{z^{2}}{2}+W_{11}^{(2)}(0)z\bar{z}
+W_{02}^{(2)}(0)\frac{\bar{z}^{2}}{2}+O(|(z,\bar{z})|^{3}),\\
\fl x_{1t}(-1)
=z{\rm exp(-\rm i\omega\tau_{\rm c})}+\bar{z}{\rm exp(\rm i\omega\tau_{\rm c})}+W_{20}^{(1)}(-1)\frac{z^{2}}{2}
+W_{11}^{(1)}(-1)z\bar{z}+W_{02}^{(1)}(-1)\frac{\bar{z}^{2}}{2}+O(|(z,\bar{z})|^{3}),\\
\fl x_{2t}(-1)
=zq_{1}{\rm exp(-\rm i\omega\tau_{\rm c})}+\bar{z}\bar{q}_{1}{\rm exp(\rm i\omega\tau_{\rm c})}+W_{20}^{(2)}(-1)\frac{z^{2}}{2}
+W_{11}^{(2)}(-1)z\bar{z}+W_{02}^{(2)}(-1)\frac{\bar{z}^{2}}{2}+O(|(z,\bar{z})|^{3}).\\
\end{eqnarray*}
Based on the definition of $f(\mu, x_{t})$, we have
\begin{equation}
g(z,\bar{z})=\bar{q}^{*}(0)f_{0}(z, \bar{z})=\bar{D}\tau_{\rm c}(1,\bar{q}_{2})\left[
\begin{array}{lll}
{f_{1}^{0}}\\
{f_{2}^{0}}\\
\end{array}\right],
\label{25} \end{equation}
where
$f_{1}^{0}=\frac{-\alpha b_{\rm m}(1-2x^{*})}{R_{\rm m}}x_{1t}(0)x_{2t}(0)-(p\beta-\frac{\alpha b_{\rm m}y^{*}}{R_{\rm m}})(x_{1t}(0))^{2}+\frac{\alpha b_{\rm m}}{R_{\rm m}}(x_{1t}(0))^{2}x_{2t}(0)$
and $f_{2}^{0}=\frac{-r}{R_{\rm m}}(x_{2t}(0))^{2}+\frac{N\alpha b_{\rm m}}{R_{\rm m}}x_{1t}(-1)x_{2t}(-1)$.

Thus,
\begin{eqnarray*}
\fl g(z,\bar{z})
=\bar{D}\tau_{\rm c}\{
\frac{-\alpha b_{\rm m}(1-2x^{*})}{R_{\rm m}}(z+\bar{z}+W_{20}^{(1)}(0)\frac{z^{2}}{2}
+W_{11}^{(1)}(0)z\bar{z}+W_{02}^{(1)}(0)\frac{\bar{z}^{2}}{2}+O(|(z,\bar{z})|^{3}))\\
\times(zq_{1}+\bar{z}\bar{q}_{1}+W_{20}^{(2)}(0)\frac{z^{2}}{2}+W_{11}^{(2)}(0)z\bar{z}
+W_{02}^{(2)}(0)\frac{\bar{z}^{2}}{2}+O(|(z,\bar{z})|^{3}))\\
-(p\beta-\frac{\alpha b_{\rm m}y^{*}}{R_{\rm m}})(z+\bar{z}+W_{20}^{(1)}(0)\frac{z^{2}}{2}
+W_{11}^{(1)}(0)z\bar{z}+W_{02}^{(1)}(0)\frac{\bar{z}^{2}}{2}+O(|(z,\bar{z})|^{3})^2)\\
+\frac{\alpha b_{\rm m}}{R_{\rm m}}(z+\bar{z}+W_{20}^{(1)}(0)\frac{z^{2}}{2}+W_{11}^{(1)}(0)z\bar{z}
+W_{02}^{(1)}(0)\frac{\bar{z}^{2}}{2}+O(|(z,\bar{z})|^{3})^2)\\
\times(zq_{1}+\bar{z}\bar{q}_{1}+W_{20}^{(2)}(0)\frac{z^{2}}{2}+W_{11}^{(2)}(0)z\bar{z}
+W_{02}^{(2)}(0)\frac{\bar{z}^{2}}{2}+O(|(z,\bar{z})|^{3}))
\}\\
+\bar{D}\tau_{\rm c}\bar{q}_{2}\{
\frac{-r}{R_{\rm m}}(zq_{1}+\bar{z}\bar{q}_{1}+W_{20}^{(2)}(0)\frac{z^{2}}{2}
+W_{11}^{(2)}(0)z\bar{z}+W_{02}^{(2)}(0)\frac{\bar{z}^{2}}{2}+O(|(z,\bar{z})|^{3}))^2\\
+\frac{N\alpha b_{\rm m}}{R_{\rm m}}(z\rm exp(-\rm i\omega\tau_{\rm c})+\bar{z}{\rm exp(\rm i\omega\tau_{\rm c})}+W_{20}^{(1)}(-1)\frac{z^{2}}{2}
+W_{11}^{(1)}(-1)z\bar{z}+W_{02}^{(1)}(-1)\frac{\bar{z}^{2}}{2}\\
+O(|(z,\bar{z})|^{3}))
\times(zq_{1}{\rm exp(-\rm i\omega\tau_{\rm c})}+\bar{z}\bar{q}_{1}{\rm exp(\rm i\omega\tau_{\rm c})}+W_{20}^{(2)}(-1)\frac{z^{2}}{2}
+W_{11}^{(2)}(-1)z\bar{z}\\
+W_{02}^{(2)}(-1)\frac{\bar{z}^{2}}{2}+O(|(z,\bar{z})|^{3}))
\},\\
\end{eqnarray*}
By comparing the coefficients with equation (\ref{24}), we obtain
\begin{eqnarray*}
\fl g_{20}
=2\bar{D}\tau_{\rm c}[
\frac{-\alpha b_{\rm m}(1-2x^{*})}{R_{\rm m}}q_{1}-(p\beta-\frac{\alpha b_{\rm m}y^{*}}{R_{\rm m}})
-\frac{r}{R_{\rm m}}\bar{q}_{2}q_{1}^2+\frac{N\alpha b_{\rm m}}{R_{\rm m}}\bar{q}_{2}q_{1}\rm exp(-\rm 2i\omega\tau_{\rm c})],\\
\fl g_{11}
=\bar{D}\tau_{\rm c}[
\frac{-\alpha b_{\rm m}(1-2x^{*})}{R_{\rm m}}(q_{1}+\bar{q}_{1})+\frac{N\alpha b_{\rm m}}{R_{\rm m}}\bar{q}_{2}(q_{1}+\bar{q}_{1})
-2(p\beta-\frac{\alpha b_{\rm m}y^{*}}{R_{\rm m}})-\frac{2r}{R_{\rm m}}\bar{q}_{2}q_{1}\bar{q}_{1}],\\
\fl g_{02}
=2\bar{D}\tau_{\rm c}[
\frac{-\alpha b_{\rm m}(1-2x^{*})}{R_{\rm m}}\bar{q}_{1}+\frac{N\alpha b_{\rm m}}{R_{\rm m}}\bar{q}_{2}\bar{q}_{1}{\rm exp(2\rm i\omega\tau_{\rm c})}
-(p\beta-\frac{\alpha b_{\rm m}y^{*}}{R_{\rm m}})-\frac{r}{R_{\rm m}}\bar{q}_{2}\bar{q}_{1}^2],\\
\end{eqnarray*}
and
\begin{eqnarray*}
\fl g_{21}
=2\bar{D}\tau_{\rm c}[\frac{-\alpha b_{\rm m}(1-2x^{*})}{R_{ \rm m}}(W_{11}^{(2)}(0)+\frac{W_{20}^{(2)}(0)}{2}
+\bar{q}_{1}\frac{W_{20}^{(1)}(0)}{2}+q_{1}W_{11}^{(1)}(0))\\
-2(p\beta-\frac{\alpha b_{\rm m}y^{*}}{R_{\rm m}})(\frac{W_{20}^{(1)}(0)}{2}+W_{11}^{(1)}(0))
+\frac{\alpha b_{\rm m}}{R_{\rm m}}(\bar{q}_{1}+2q_{1})\\
-\frac{r}{R_{\rm m}}\bar{q}_{2}(\bar{q}_{1}W_{20}^{(2)}(0)+2q_{1}W_{11}^{(2)}(0))
+\frac{N\alpha b_{\rm m}}{R_{\rm m}}\bar{q}_{2}(W_{11}^{(2)}(-1){\rm exp(-\rm i\omega\tau_{\rm c})}\\
+\frac{W_{20}^{(2)}(-1)}{2}{\rm exp(\rm i\omega\tau_{\rm c})}+\bar{q}_{1}\frac{W_{20}^{(1)}(-1)}{2}{\rm exp(\rm i\omega\tau_{\rm c})}
+q_{1}W_{11}^{(1)}(-1){\rm exp(-\rm i\omega\tau_{\rm c})})].\\
\end{eqnarray*}
To determine $g_{21}$, we need to compute $W_{20}(\theta)$ and $W_{11}(\theta)$.
Based on equations (\ref{21}) and (\ref{24}), we have
\begin{eqnarray}
\dot{W}
&=&\dot{x}_{t}-\dot{z}q_{1}+\dot{\bar{zq_{1}}}\nonumber\\
&=&AW+H(z, \bar{z}, \theta),\nonumber\\
\label{26}
\end{eqnarray}
where
\begin{equation}
H(z, \bar{z}, \theta)=H_{20}(\theta)\frac{z^{2}}{2}+H_{11}(\theta)z\bar{z}+H_{02}(\theta)\frac{\bar{z}^{2}}{2}+....
\label{27} \end{equation}
Note that on the center manifold $C_{0}$ near to the origin,
we have
\begin{equation}
\dot{W}=W_{z}\dot{z}+W_{\bar{z}}\dot{\bar{z}}.
\label{28} \end{equation}
Thus, we obtain
\begin{equation}
(A-2\rm i\omega\tau_{\rm c})W_{20}(\theta)=-H_{20}(\theta)
\label{29} \end{equation}
and
\begin{equation}
AW_{11}(\theta)=-H_{11}(\theta).
\label{30} \end{equation}
By using equation (\ref{26}), for $\theta\in[-1, 0)$ we have
\begin{equation}
H(z, \bar{z}, \theta)=-\bar{q}^{*}f_{0}q(\theta)-q^{*}f_{0}(0)\bar{q}(\theta)=-gq(\theta)-\bar{g}\bar{q}(\theta).
\label{31} \end{equation}
Comparing the coefficients with equation (\ref{27}), we obtain
\begin{equation}
H_{20} (\theta)=-g_{20}q(\theta)-\bar{g}_{02}\bar{q}(\theta)
\label{32} \end{equation}
and
\begin{equation}
 H_{11} (\theta)=-g_{11}q(\theta)-\bar{g}_{11}\bar{q}(\theta).
\label{33} \end{equation}
From equations (\ref{29}), (\ref{30}), (\ref{32}), (\ref{33}), and the  definition of $A$, we get
$$\dot{W}_{20}(\theta)=2{\rm i}\omega\tau_{\rm c} W_{20}(\theta)+g_{20}q(\theta)+\bar{g}_{02}\bar{q}(\theta).$$
Noticing $q(\theta)=q(0){\rm exp({\rm i}\omega\tau_{\rm c}\theta)}$, we have
\begin{equation}
 W_{20} (\theta)=\frac{ig_{20}q(0){\rm exp(\rm i\omega\tau_{\rm c}\theta)}}{\omega\tau_{\rm c}}+\frac{{\rm i}\bar{g}_{02}\bar{q}(0){\rm exp(-{\rm i}\omega\tau_{\rm c}\theta)}}
{3\omega\tau_{\rm c}}+E_{1}{\rm exp(2\rm i\omega\tau_{\rm c}\theta)},
\label{34} \end{equation}
where $E_{1}=(E_{1}^{1}, E_{1}^{2})\in R^{2}$ is a constant vector.
Similarly, we have
\begin{equation}
W_{11} (\theta)=-\frac{{\rm i}g_{11}q(0){\rm exp({\rm i}\omega\tau_{\rm c}\theta)}}{\omega\tau_{\rm c}}+\frac{{\rm i}\bar{g}_{11}\bar{q}(0){\rm exp(-\rm i\omega\tau_{ \rm c}\theta)}}{\omega\tau_{\rm c}}+E_{2},
\label{35} \end{equation}
where $E_{2}=(E_{2}^{1}, E_{2}^{2})\in R^{2}$ is a constant vector.
Now we will try to find $E_{1}$ and $E_{2}$. From the definition of $A$, equations (\ref{29}) and (\ref{30}), we obtain
\begin{equation}
\int_{-1}^{0} d\eta(\theta)W_{20} (\theta)=2\rm i\omega\tau_{\rm c} W_{20}(0)-H_{20}(0)
\label{36} \end{equation}
and
\begin{equation}
\int_{-1}^{0} d\eta(\theta)W_{11} (\theta)=-H_{11}(0),
\label{37} \end{equation}
where $d\eta(\theta)=\eta(\theta,0)$.

Based on equations (\ref{26}) and (\ref{27}), we have
\begin{eqnarray}
\fl H_{20} (0)=-g_{20}q(0)-\bar{g}_{02}\bar{q}(0)+2\tau_{\rm c}
\left[
\begin{array}{lll}
\frac{-\alpha b_{\rm m}(1-2x^{*})}{R_{\rm m}}q_{1}-(p\beta-\frac{\alpha b_{\rm m}y^{*}}{R_{\rm m}})\\
-\frac{r}{R_{\rm m}}q_{1}^2+\frac{N\alpha b_{\rm m}}{R_{\rm m}}q_{1}{\rm exp(-2\rm i\omega\tau_{\rm c}\theta)}\\
\end{array}\right]
\label{38}
\end{eqnarray}
and
\begin{eqnarray}
\fl H_{11} (0)=-g_{11}q(0)-\bar{g}_{11}\bar{q}(0)+
\tau_{\rm c}\left[
\begin{array}{l}
\frac{-\alpha b_{\rm m}(1-2x^{*})}{R_{\rm m}}(q_{1}+\bar{q}_{1})-2(p\beta-\frac{\alpha b_{\rm m}y^{*}}{R_{\rm m}})\\
-\frac{2r}{R_{\rm m}}q_{1}\bar{q}_{1}+\frac{N\alpha b_{\rm m}}{R_{\rm m}}(q_{1}+\bar{q}_{1})\\
\end{array}\right].
\label{39}
\end{eqnarray}
Substituting equations (\ref{36}) and (\ref{38}) into equation (\ref{34}) and noticing that
$$(\rm i\omega\tau_{\rm c} I-\int_{-1}^{0}{\rm exp(\rm i\omega\tau_{\rm c}\theta)}d\eta(\theta))q(0)=0$$
and
$$(-\rm i\omega\tau_{\rm c} I-\int_{-1}^{0}{\rm exp(-\rm i\omega\tau_{\rm c}\theta)}d\eta(\theta))\bar{q}(0)=0,$$
we obtain
\begin{eqnarray*}
(2\rm i\omega\tau_{\rm c} I-\int_{-1}^{0}{\rm exp(2{\rm i}\omega\tau_{\rm c}\theta)}d\eta(\theta))E_{1}
=2\tau_{\rm c}
\left[\begin{array}{lll}
\frac{-\alpha b_{ \rm m}(1-2x^{*})}{R_{\rm m}}q_{1}-(p\beta-\frac{\alpha b_{\rm m}y^{*}}{R_{\rm m}})\\
-\frac{r}{R_{\rm m}}q_{1}^2+\frac{N\alpha b_{\rm m}}{R_{\rm m}}q_{1}{\rm exp(-2\rm i\omega\tau_{\rm c})}\\
\end{array}\right],\\
\end{eqnarray*}
which is
\begin{eqnarray*}
\fl\left[
\begin{array}{lll}
2{\rm i}\omega-a_{11} & -a_{12} \\
-b_{21}{\rm exp(-2{\rm i}\omega\tau_{\rm c})} & 2{\rm i}\omega-a_{22}-b_{22}{\rm exp(-2\rm i\omega\tau_{\rm c})}\\
\end{array}\right]
E_{1}
=2
\left[\begin{array}{lll}
\frac{-\alpha b_{\rm m}(1-2x^{*})}{R_{\rm m}}q_{1}-(p\beta-\frac{\alpha b_{\rm m}y^{*}}{R_{\rm m}})\\
-\frac{r}{R_{\rm m}}q_{1}^2+\frac{N\alpha b_{\rm m}}{R_{\rm m}}q_{1}{\rm exp(-2\rm i\omega\tau_{\rm c})}\\
 \end{array}\right].\\
\end{eqnarray*}
We further get
\begin{eqnarray*}
E_{1}^{1}=\frac{2}{A_{1}}
\left|\begin{array}{cccc}
\frac{-\alpha b_{\rm m}(1-2x^{*})q_{1}}{R_{\rm m}}-p\beta+\frac{\alpha b_{\rm m}y^{*}}{R_{\rm m}}&-a_{12}\\
-\frac{rq_{1}^2}{R_{\rm m}}+\frac{N\alpha b_{\rm m}q_{1}}{R_{\rm m}}{\rm exp(-2{\rm i}\omega\tau_{\rm c})}&2\rm i\omega-a_{22}-b_{22}{\rm exp(-2\rm i\omega\tau_{\rm c})}\\
 \end{array}\right|
\end{eqnarray*}
and
\begin{eqnarray*}
E_{1}^{2}=\frac{2}{A_{1}}
\left|\begin{array}{cccc}
2{\rm i}\omega-a_{11} &\frac{-\alpha b_{\rm m}(1-2x^{*})}{R_{\rm m}}q_{1}-(p\beta-\frac{\alpha b_{\rm m}y^{*}}{R_{\rm m}})\\
-b_{21}{\rm exp(-2{\rm i}\omega\tau_{\rm c})}&-\frac{r}{R_{\rm m}}q_{1}^2+\frac{N\alpha b_{\rm m}}{R_{\rm m}}q_{1}{\rm exp(-2{\rm i}\omega\tau_{\rm c})}\\
\end{array}\right|,
\end{eqnarray*}
where
\begin{eqnarray*}
A_{1}=
\left|\begin{array}{cccc}
    2{\rm i}\omega-a_{11}&-a_{12}\\
-b_{21}{\rm exp(-2{\rm i}\omega\tau_{\rm c})}&2{\rm i}\omega-a_{22}-b_{22}{\rm exp(-2{\rm i}\omega\tau_{\rm c})}\\
\end{array}\right|.
\end{eqnarray*}

Similarly, substituting equations (\ref{37}) and (\ref{39}) into (\ref{35}), we obtain
\begin{eqnarray*}
\left[
\begin{array}{lll}
a_{11} & a_{12} \\
b_{21} & a_{22}+b_{22}\\
\end{array}\right]
E_{2}=
\left[
\begin{array}{lll}
\frac{\alpha b_{\rm m}(1-2x^{*})}{R_{\rm m}}(q_{1}+\bar{q}_{1})+2(p\beta-\frac{\alpha b_{\rm m}y^{*}}{R_{\rm m}})\\
\frac{2r}{R_{\rm m}}q_{1}\bar{q}_{1}-\frac{N\alpha b_{\rm m}}{R_{\rm m}}(q_{1}+\bar{q}_{1})\\
\end{array}\right].
\end{eqnarray*}

Therefore we can obtain
\begin{eqnarray*}
E_{2}^{1}=\frac{1}{A_{2}}
\left|\begin{array}{cccc}
\frac{\alpha b_{\rm m}(1-2x^{*})}{R_{\rm m}}(q_{1}+\bar{q}_{1})+2(p\beta-\frac{\alpha b_{\rm m}y^{*}}{R_{\rm m}})&a_{12}\\
\frac{2r}{R_{\rm m}}q_{1}\bar{q}_{1}-\frac{N\alpha b_{\rm m}}{R_{\rm m}}(q_{1}+\bar{q}_{1})&a_{22}+b_{22}\\
 \end{array}\right|
\end{eqnarray*}
and
\begin{eqnarray*}
E_{2}^{2}=\frac{1}{A_{2}}
\left|\begin{array}{cccc}
a_{11} &\frac{\alpha b_{\rm m}(1-2x^{*})}{R_{\rm m}}(q_{1}+\bar{q}_{1})+2(p\beta-\frac{\alpha b_{\rm m}y^{*}}{R_{\rm m}})\\
b_{21}&\frac{2r}{R_{\rm m}}q_{1}\bar{q}_{1}-\frac{N\alpha b_{\rm m}}{R_{\rm m}}(q_{1}+\bar{q}_{1})\\
\end{array}\right|,
\end{eqnarray*}

where
\begin{eqnarray*}
A_{2}=
\left|\begin{array}{cccc}
    a_{11}&a_{12}\\
b_{21}&a_{22}+b_{22}\\
\end{array}\right|.
\end{eqnarray*}

Thus, we can compute $W_{20}(\theta)$ and $W_{11}(\theta)$ from equations (\ref{34}) and (\ref{35}) and determine the following values to investigate the qualities of
bifurcation periodic solution in the center manifold at the critical value $\tau_{\rm c}$.
And then we can evaluate the following values
\begin{eqnarray}
&{}&c_{1}(0)=\frac{{\rm i}(g_{20}g_{11}-2|g_{11}|^{2}-\frac{|g_{02}|^{2}}{3})}{2\omega\tau_{\rm c}}+\frac{g_{21}}{2},\nonumber\\
&{}&\mu_{2}=-\frac{Re{c_{1}(0)}}{Re{\lambda^{'}}(\tau_{\rm c})},\nonumber\\
\label{40}
\end{eqnarray}
and
\begin{eqnarray}
&{}&\beta_{2}=2Re{c_{1}(0)},\nonumber\\
\label{41}
\end{eqnarray}
which are the quantities for determining of bifurcating periodic solutions in the center manifold at $\tau_{\rm c}$.
Specifically,  $\mu_{2}$ determines the direction of Hopf bifurcation: if $\mu_{2}>0$, then the Hopf bifurcation is
supercritical and the bifurcating periodic solution exists for $\tau>\tau_{\rm c}$; if $\mu_{2}<0$, then the Hopf bifurcation is subcritical and the bifurcating periodic solution exists for $\tau<\tau_{\rm c}$. The parameter $\beta_{2}$ determines the stability of the bifurcating periodic solution: bifurcating periodic solutions are stable if $\beta_{2}<0$ and unstable if $\beta_{2}>0$.

In the case of $e_{C}<r<e_{D}$ and $\alpha b_{\rm m}(1-\frac{e_{C}}{r})<p\beta$,
the equation system has three fixed points in the parameter space of $0\leq x\leq1$ and $y \geq 0$.
They are $F_{0}$, $F_{1}$, and $F_{3}$, respectively. We know that the characteristic equation of equation (\ref{2}) at the fixed point $F_{0}$,
which can be written as
\begin{equation}
\lambda^{2}-(r-e_{D}+p\beta)\lambda+p\beta(r-e_{D})=0.
\label{42} \end{equation}
Since $H<0$, $c^{2}-d^{2}>0$, and the eigenvalues for $\tau=0$ are $\lambda_{1}=r-e_{D}<0$ and $\lambda_{2}=p\beta>0$,
respectively, $F_{0}$ is unstable for $\tau\geq0$ according to Proposition 1 and reference~\cite{kuang_93}.

The characteristic equation of equation (\ref{2}) at the fixed point $F_{1}$ is
\begin{equation}
\lambda^{2}-(r-e_{D}-p\beta)\lambda-\alpha e_{D} \lambda {\rm e^{-\lambda\tau}}
-p\beta(r-e_{D})-p\beta\alpha e_{D}{\rm e^{-\lambda\tau}} =0.
\label{43} \end{equation}
The eigenvalues for $\tau=0$ are $\lambda_{1}=-p\beta<0$ and $\lambda_{2}=r-e_{C}>0$, respectively,
which means that the fixed point $F_{1}$ for $\tau=0$ is unstable. Moreover, since $c^{2}-d^{2}<0$, according to Proposition 1 we know that there exist a pair of purely imaginary solutions which are $\lambda=\pm \rm i\omega_{+}$ with $\omega_{+}>0$ and
${\frac{d(R_{e}\lambda)}{d\tau}}|_{\lambda=\rm i\omega_{+}}>0$.
Therefore, according to reference~\cite{kuang_93}, the unstable fixed point $F_{1}$ for $\tau=0$ never becomes stable for $\tau>0$, that is to say,
$F_{1}$ remains unstable for $\tau\geq0$.

The characteristic equation of equation (\ref{2}) at the fixed point $F_{3}$ is
\begin{equation}
\lambda^{2}-(L+M)\lambda- \alpha e_{D}\lambda e^{-\lambda\tau}{\rm e^{-\lambda\tau}}
+ML+M\alpha e_{D}{\rm e^{-\lambda\tau}} =0,
\label{44} \end{equation}
where $L=e_{D}-r-2\alpha e_{D}$ and $M=\alpha b_{\rm m}(1-\frac{e_{C}}{r})-p\beta$. Since $H<0$ and $c^{2}-d^{2}>0$, and the eigenvalues for $\tau=0$ are $\lambda_{1}=\alpha b_{\rm m}(1-\frac{e_{C}}{r})-p\beta<0$ and $\lambda_{2}=e_{C}-r<0$, respectively, $F_{3}$ is asymptotically stable $\tau\geq0$ according to Proposition 1 and reference~\cite{kuang_93}.

In the special case of $p\beta=\alpha b_{\rm m}(1-\frac{e_{C}}{r})$, we find that there are three fixed points, which are $F_{0}$, $F_{1}$ and $F_{3}$, respectively. The characteristic equation of equation (\ref{2}) at the fixed point $F_{0}$ can be written as
\begin{equation}
\lambda^{2}-(r-e_{D}+p\beta)\lambda+p\beta(r-e_{D})=0.
\label{45} \end{equation}
Since $H<0$, $c^{2}-d^{2}>0$, and the eigenvalues are $\lambda_{1}=r-e_{D}<0$ and $\lambda_{2}=p\beta>0$, according to Proposition 1 and reference~\cite{kuang_93} $F_{0}$ is unstable for $\tau\geq0$.

The characteristic equation of equation (\ref{2}) at the fixed point $F_{1}$ is
\begin{equation}
\lambda^{2}-(r-e_{D}-p\beta)\lambda-\alpha e_{D} \lambda {\rm e^{-\lambda\tau}}
-p\beta(r-e_{D})-p\beta\alpha e_{D}{\rm e^{-\lambda\tau}} =0.
\label{46} \end{equation}
The eigenvalues for $\tau=0$ are $\lambda_{1}=-p\beta<0$ and $\lambda_{2}=r-e_{C}>0$, meaning that the fixed point $F_{1}$ for $\tau=0$ is unstable.
Moreover, since $c^{2}-d^{2}<0$, and there exists a pair of purely imaginary solutions which are $\lambda=\pm \rm i\omega_{+}$ with $\omega_{+}>0$.
We further obtain ${\frac{d(R_{e}\lambda)}{d\tau}}|_{\lambda=\rm i\omega_{+}}>0$, hence $F_{1}$ remains unstable for $\tau\geq0$~\cite{kuang_93}.

The characteristic equation of equation (\ref{2}) at the fixed point $F_{3}$ is
\begin{equation}
\lambda^{2}-(L+M)\lambda- \alpha e_{D}\lambda {\rm e^{-\lambda\tau}}
+ML+M\alpha e_{D}{\rm e^{-\lambda\tau}} =0,
\label{47} \end{equation}
where $a=r-e_{D}(1-2\alpha)>0$, $b=-\alpha e_{D}$, $c=0$, and $d=0$. Since $c+d=0$ and $a^{2}-b^{2}=(r-e_{C})(r-e_{D}(1-3\alpha))>0$, $F_{3}$ is stable, but not asymptotically stable for $\tau\geq0$~\cite{kuang_93}.

Next, we provide the theoretical analysis of the equilibrium points for the special case
of $r=e_{D}$.
In this case, we have $R_{\rm m}-\frac{Nb_{\rm m}}{r}=0$. In dependence of the efficiency of inspection
and punishment,  we can further distinguish three following sub-cases. Firstly, if the product of $p\beta$ exceeds $b_{\rm m}(1-\frac{e_{C}}{r})$, we then have $K>1$. As a result, the equation system has three fixed points in the parameter space. They are $F_{0}$, $F_{1}$,  and  $F_{3}$, respectively. The characteristic equation of equation (\ref{2}) at the fixed point $F_{0}$,
which can be written as
\begin{equation}
\lambda^{2}-p\beta\lambda=0.
\label{48} \end{equation}
Since $H<0$, $c^{2}-d^{2}>0$, and the eigenvalues for $\tau=0$ are $\lambda_{1}=0$ and $\lambda_{2}=p\beta>0$, according to Proposition 1 and reference~\cite{kuang_93} $F_{0}$ is unstable for $\tau\geq0$.

The characteristic equation of equation (\ref{2}) at the fixed point $F_{1}$ is
\begin{equation}
\lambda^{2}+p\beta\lambda-\alpha e_{ D} \lambda {\rm e^{-\lambda\tau}}-p\beta\alpha e_{D}{\rm e^{-\lambda\tau}} =0.
\label{49} \end{equation}
The eigenvalues for $\tau=0$ are $\lambda_{1}=-p\beta<0$ and $\lambda_{2}=r-e_{C}>0$, hence $F_{1}$ for $\tau=0$ is unstable. Moreover, since $c^{2}-d^{2}<0$, the pair of purely imaginary solutions are $\lambda=\pm \rm i\omega_{+}$ with $\omega_{+}>0$. We further obtain ${\frac{d(R_{e}\lambda)}{d\tau}}|_{\lambda=\rm i\omega_{+}}>0$, hence $F_{1}$ remains unstable for $\tau\geq0$~\cite{kuang_93}.

The characteristic equation of equation (\ref{2}) at the fixed point $F_{3}$ is
\begin{equation}
\lambda^{2}-(L+M)\lambda- \alpha e_{D}\lambda {\rm e^{-\lambda\tau}}
+ML+M\alpha e_{D}{\rm e^{-\lambda\tau}} =0,
\label{50} \end{equation}
where $L=e_{D}-r-2\alpha e_{D}$ and $M=\alpha b_{\rm m}(1-\frac{e_{C}}{r})-p\beta$.
Since $H<0$ and $c^{2}-d^{2}>0$, there does not exist a purely imaginary solution and stability does not change for any $\tau\geq0$. Moreover, the eigenvalues for $\tau=0$ are $\lambda_{1}=\alpha b_{\rm m}(1-\frac{e_{C}}{r})-p\beta<0$ and $\lambda_{2}=e_{C}-r<0$, hence $F_{3}$ is asymptotically stable for $\tau\geq0$~\cite{kuang_93}.

Secondly, when the above mentioned institutions are less effective, the term $\alpha b_{\rm m}(1-\frac{e_{C}}{r})$ exceeds $p\beta$. We then have $K>1$. As a result, the equation system has four fixed points, which are  $F_{0}$, $F_{1}$, $F_{3}$, and $F_{4}$. The characteristic equation of equation (\ref{2}) at the fixed point $F_{0}$ is
\begin{equation}
\lambda^{2}-p\beta\lambda=0.
\label{51} \end{equation}
Since $H<0$, $c^{2}-d^{2}>0$, and the eigenvalues are $\lambda_{1}=0$ and $\lambda_{2}=p\beta>0$, $F_{0}$ is unstable for $\tau\geq0$ according to Proposition 1 and reference~\cite{kuang_93}.

The characteristic equation of equation (\ref{2}) at the fixed point $F_{1}$ is
\begin{equation}
\lambda^{2}+p\beta\lambda-\alpha e_{D} \lambda {\rm e^{-\lambda\tau}}-p\beta\alpha e_{D}{\rm e^{-\lambda\tau}} =0.
\label{52} \end{equation}
These eigenvalues of the fixed point $F_{1}$
for $\tau=0$ are $\lambda_{1}=-p\beta<0$ and  $\lambda_{2}=r-e_{C}>0$, respectively, which means that the fixed point $F_{1}$ for $\tau=0$ is unstable.
Moreover, since $c^{2}-d^{2}<0$, according to Proposition 1 we know that there exist a pair of purely imaginary solutions
which are $\lambda=\pm \rm i\omega_{+}$ with $\omega_{+}>0$.
We further obtain  ${\frac{d(R_{e}\lambda)}{d\tau}}|_{\lambda=\rm i\omega_{+}}>0$, therefore $F_{1}$ remains unstable for $\tau\geq0$~\cite{kuang_93}.

The characteristic equation of equation (\ref{2}) at the fixed point $F_{3}$ is
\begin{equation}
\lambda^{2}-(L+M)\lambda- \alpha e_{D}\lambda {\rm e^{-\lambda\tau}}
+ML+M\alpha e_{D}{\rm e^{-\lambda\tau}} =0,
\label{53} \end{equation}
where $L=e_{D}-r-2\alpha e_{D}$ and $M=\alpha b_{\rm m}(1-\frac{e_{C}}{r})-p\beta$.
Since $H<0$ and $c^{2}-d^{2}>0$, and the eigenvalues are $\lambda_{1}=\alpha b_{\rm m}(1-\frac{e_{C}}{r})-p\beta>0$ and $\lambda_{2}=e_{C}-r<0$, $F_{3}$ is unstable $\tau\geq0$ according to Proposition 1 and reference~\cite{kuang_93}.

For the fixed point $F_{4}$ the characteristic equation can be written as
\begin{equation}
\lambda^{2}+\frac{2Np\beta}{\alpha R_{\rm m}}\lambda-\frac{Np\beta}{\alpha R_{\rm m}}\lambda {\rm e^{-\lambda\tau}}
+Np\beta K(1-K)\frac{\alpha b_{\rm m}}{R_{\rm m}}{\rm e^{-\lambda\tau}}=0,
\label{54} \end{equation}
where $K=\frac{1}{\alpha}-\frac{rR_{\rm m}}{\alpha b_{\rm m}N}+\frac{p\beta rR_{\rm m}}{N\alpha^{2}b_{ \rm m}^{2}}=\frac{p\beta}{\alpha^2b_{\rm m}}$.
Since these eigenvalues of the fixed point $F_{4}$ for $\tau=0$ satisfies
$\lambda_{1}+\lambda_{2}<0$ and $\lambda_{1}\lambda_{2}>0$, $F_{4}$ is asymptotically stable for $\tau=0$.
Moreover, since $c^{2}<d^{2}$, equation (\ref{8}) has only one pair of purely imaginary solutions and $\lambda=\pm \rm i\omega_{+}$ with $\omega_{+}>0$,
where
\begin{equation}
\omega_{+}^{2}=\frac{1}{2}[-\frac{3N^2p^2\beta^2}{\alpha^2 R^2_{\rm m}}+
\sqrt{(\frac{3N^2p^2\beta^2}{\alpha^2 R^2_{\rm m}})^{2}+4Np\beta K
(1-K)\frac{\alpha b_{\rm m}}{R_{\rm m}}}].
\label{55} \end{equation}
According to reference~\cite{kuang_93} there exists a critical time delay $\tau_{\rm c}=\frac{\theta_{1}}{\omega_{+}}$, where
$\theta_{1}$ satisfies
\begin{equation}
\cos\theta_{1}=-\frac{(ab^{2}-d)\omega_{+}^{2}}{b^{2}\omega_{+}^{2}+d^{2}} \hspace{0.5cm}
%\label{56} \end{equation}
\textrm{and} \hspace{0.5cm}
%\begin{equation}
\sin\theta_{1}=-\frac{ad^{2}\omega_{+}+b\omega_{+}^{3}}{b^{2}\omega_{+}^{2}+d^{2}}.
\label{57} \end{equation}
Hence $F_{4}$ is  asymptotically stable for $\tau<\tau_{\rm c}$ and unstable for $\tau>\tau_{\rm c}$.
For $\tau=\tau_{\rm c}$, there exists a Hopf bifurcation in the system.
Furthermore, we can determine the direction of Hopf bifurcation and the stability of the bifurcating periodic solutions by the analysis mentioned before.

Thirdly, for $p\beta=\alpha b_{\rm m}(1-\frac{e_{C}}{r})$,
we  have $K=1$
and $R_{\rm m}-\frac{Nb_{\rm m}(1-\alpha)}{r}=\frac{p\beta R_{\rm m}}{\alpha b_{\rm m}}$.
As a result, the system has three fixed points in the parameter space: $F_{0}$, $F_{1}$, and  $F_{3}$.
The characteristic equation of equation~(\ref{2}) at the fixed point $F_{0}$ is
\begin{equation}
\lambda^{2}-p\beta\lambda=0.
\label{58} \end{equation}
Since $H<0$, $c^{2}-d^{2}>0$, and the eigenvalues
are $\lambda_{1}=0$ and $\lambda_{2}=p\beta>0$, according to Proposition 1 and reference~\cite{kuang_93} $F_{0}$ is unstable for $\tau\geq0$.

The characteristic equation of equation (\ref{2}) at the fixed point $F_{1}$ is
\begin{equation}
%\begin{aligned}
\lambda^{2}+p\beta\lambda-\alpha e_{D} \lambda {\rm e^{-\lambda\tau}}-p\beta\alpha e_{D}{\rm e^{-\lambda\tau}} =0.
%\end{aligned}
\label{59} \end{equation}
The eigenvalues are $\lambda_{1}=-p\beta<0$ and  $\lambda_{2}=r-e_{C}>0$, yielding $F_{1}$ is unstable for $\tau=0$. Moreover, since $c^{2}-d^{2}<0$, there exist a pair of purely imaginary solutions which are $\lambda=\pm \rm i\omega_{+}$ with $\omega_{+}>0$. We further obtain  ${\frac{d(R_{e}\lambda)}{d\tau}}|_{\lambda=\rm i\omega_{+}}>0$. Therefore, $F_{1}$ remains unstable for $\tau\geq0$~\cite{kuang_93}.

The characteristic equation of the system at the fixed point $F_{3}$ is
\begin{equation}
\lambda^{2}-(L+M)\lambda- \alpha e_{D}\lambda {\rm e^{-\lambda\tau}}
+ML+M\alpha e_{D}{\rm e^{-\lambda\tau}} =0,
\label{60} \end{equation}
where $a=\alpha e_{D}>0$, $b=-\alpha e_{D}$, $c=0$, and $d=0$. Since $c+d=0$ and $a^{2}-b^{2}=3(r-e_{C})e_{D}\alpha>0$, $F_{3}$ is stable, but not asymptotically stable~\cite{kuang_93}.\\

\subsection*{\rm{\textbf{B3. Rapidly growing resource pool}}}

In the case of $r>e_{D}$ and $p\beta>\alpha b_{\rm m}(1-\frac{e_{C}}{r})$,
the equation system has four fixed points which are $F_{0}$, $F_{1}$, $F_{2}$, and $F_{3}$, respectively. The characteristic equation of equation (\ref{2}) at the fixed point $F_{0}$ is
\begin{equation}
\lambda^{2}-(r-e_{D}+p\beta)\lambda+p\beta(r-e_{D})=0\,.
\label{61} \end{equation}
Since $H<0$, $c^{2}-d^{2}>0$, and the eigenvalues
are $\lambda_{1}=r-e_{D}>0$ and $\lambda_{2}=p\beta>0$, $F_{0}$ is unstable for $\tau\geq0$ according to Proposition 1 and reference~\cite{kuang_93}.

The characteristic equation of equation (\ref{2}) at the fixed point $F_{1}$ is
\begin{equation}
\lambda^{2}-(r-e_{D}-p\beta)\lambda-\alpha e_{D} \lambda {\rm e^{-\lambda\tau}}
-p\beta(r-e_{D})-p\beta\alpha e_{D}{\rm e^{-\lambda\tau}} =0.
\label{62} \end{equation}
The related eigenvalues are $\lambda_{1}=-p\beta<0$ and $\lambda_{2}=r-e_{C}>0$, hence $F_{1}$ is unstable for $\tau=0$.
Moreover, since $c^{2}<d^{2}$ for $e_{D}<r<e_{D}(1+\alpha)$,
there exist a pair of purely imaginary solutions
which are $\lambda=\pm \rm i\omega_{+}$ with $\omega_{+}>0$ according to Proposition 1.
We further obtain ${\frac{d(R_{e}\lambda)}{d\tau}}|_{\lambda=\rm i\omega_{+}}>0$, therefore $F_{1}$ remains unstable for $\tau\geq0$ ~\cite{kuang_93}. While for $r>e_{D}(1+\alpha)$, since $H<0$ and $c^{2}>d^{2}$,
there does not exist a pair of purely imaginary solutions and there are no stability switches for $\tau\geq0$ according to Proposition 1.
Therefore, the stability of the fixed point $F_{1}$ for $\tau>0$ is the same with $\tau=0$, and $F_{1}$ remains unstable for any $\tau\geq0$~\cite{kuang_93}.

The characteristic equation of equation (\ref{2}) at the fixed point $F_{2}$ is
\begin{equation}
\lambda^{2}-(e_{D}-r+p\beta-\alpha b_{\rm m}+\alpha b_{\rm m}e_{D})\lambda
+(p\beta-\alpha b_{\rm m}+\alpha b_{\rm m}e_{D})(r-e_{D})=0.
\label{63} \end{equation}
Since $H<0$, $c^{2}-d^{2}>0$, and the eigenvalues are  $\lambda_{1}=p\beta-\alpha b_{\rm m}+\alpha b_{\rm m}\frac{e_{D}}{r}>0$ and $\lambda_{2}=e_{D}-r<0$, according to Proposition 1 and reference~\cite{kuang_93} $F_{2}$ is unstable for $\tau\geq0$.

The characteristic equation of the system at the fixed point $F_{3}$ is
\begin{equation}
\lambda^{2}-(L+M)\lambda- \alpha e_{D}\lambda {\rm e^{-\lambda\tau}}
+ML+M\alpha e_{D}{\rm e^{-\lambda\tau}} =0.
\label{64} \end{equation}
Since $H<0$, $c^{2}-d^{2}>0$, and the eigenvalues are
$\lambda_{1}=\alpha b_{\rm m}(1-\frac{e_{C}}{r})-p\beta<0$
and $\lambda_{2}=e_{C}-r<0$, $F_{3}$ is asymptotically  stable for $\tau\geq0$ according to Proposition 1 and reference~\cite{kuang_93}.

For $r>e_{D}$ and $\alpha b_{\rm m}(1-\frac{e_{D}}{r})< p\beta< \alpha b_{\rm m}(1-\frac{e_{C}}{r})$,
the equation system has five fixed points, which are $F_{0}$, $F_{1}$, $F_{2}$, $F_{3}$ and $F_{4}$, respectively. The characteristic equation of equation (\ref{2}) at the fixed point $F_{0}$ is
\begin{equation}
\lambda^{2}-(r-e_{D}+p\beta)\lambda+p\beta(r-e_{D})=0.
\label{65} \end{equation}
Since $H<0$, $c^{2}-d^{2}>0$, and the eigenvalues are $\lambda_{1}=r-e_{D}>0$ and $\lambda_{2}=p\beta>0$, according to Proposition 1 and reference~\cite{kuang_93} $F_{0}$ is unstable for $\tau\geq0$.

The characteristic equation of equation (\ref{2}) at the fixed point $F_{1}$ is
\begin{equation}
\lambda^{2}-(r-e_{D}-p\beta)\lambda-\alpha e_{D} \lambda {\rm e^{-\lambda\tau}}
-p\beta(r-e_{D})-p\beta\alpha e_{D}{\rm e^{-\lambda\tau}} =0.
\label{66} \end{equation}
The related eigenvalues are $\lambda_{1}=-p\beta<0$ and $\lambda_{2}=r-e_{C}>0$, yielding that $F_{1}$ is unstable for $\tau=0$. Moreover, since $c^{2}<d^{2}$ for $e_{D}<r<e_{D}(1+\alpha)$, there exist a pair of purely imaginary solutions,
which are $\lambda=\pm \rm i\omega_{+}$ with $\omega_{+}>0$.
We further obtain ${\frac{d(R_{e}\lambda)}{d\tau}}|_{\lambda=\rm i\omega_{+}}>0$, therefore $F_{1}$ remains unstable for $\tau\geq0$~\cite{kuang_93}. While for $r>e_{D}(1+\alpha)$, since $H<0$ and $c^{2}>d^{2}$, there do not exist a pair of purely imaginary solutions and there are no stability switches for any $\tau\geq0$ according to Proposition 1.
Therefore, the stability of the fixed point $F_{1}$ for $\tau>0$ is the same with $\tau=0$, and $F_{1}$ remains unstable for $\tau\geq0$~\cite{kuang_93}.

The characteristic equation of equation (\ref{2}) at the fixed point $F_{2}$ is
\begin{equation}
 \lambda^{2}-(e_{D}-r+p\beta-\alpha b_{\rm m}+\alpha b_{\rm m}e_{D})\lambda
+(p\beta-\alpha b_{\rm m}+\alpha b_{\rm m}e_{D})(r-e_{D})=0.
\label{67} \end{equation}
Since $H<0$, $c^{2}-d^{2}>0$, and the eigenvalues are $\lambda_{1}=p\beta-\alpha b_{\rm m}+\alpha b_{\rm m}\frac{e_{D}}{r}>0$ and $\lambda_{2}=e_{D}-r<0$, $F_{2}$ is unstable for $\tau\geq0$ according to Proposition 1 and reference~\cite{kuang_93}.

The characteristic equation of the system at the fixed point $F_{3}$ is
\begin{equation}
\lambda^{2}-(L+M)\lambda- \alpha e_{D}\lambda {\rm e^{-\lambda\tau}}
+ML+M\alpha e_{D}{\rm e^{-\lambda\tau}} =0.
\label{68} \end{equation}
Since $H<0$, and $c^{2}-d^{2}>0$, and the eigenvalues are
$\lambda_{1}=\alpha b_{\rm m}(1-\frac{e_{C}}{r})-p\beta>0$
and $\lambda_{2}=e_{C}-r<0$, according to Proposition 1 and reference~\cite{kuang_93} $F_{3}$ is unstable for $\tau\geq0$.

The characteristic equation for the fixed point $F_{4}$ is
\begin{equation}
\fl\lambda^{2}+(\frac{2rp\beta}{\alpha b_{\rm m}}+e_{D}-r)\lambda+(r-\frac{rp\beta}{\alpha b_{\rm m}}-e_{D})\lambda {\rm e^{-\lambda\tau}}
+Np\beta K(1-K)\frac{\alpha b_{\rm m}}{R_{\rm m}}{\rm e^{-\lambda\tau}}=0,
\label{69} \end{equation}
where $K=\frac{1}{\alpha}-\frac{rR_{\rm m}}{\alpha b_{\rm m}N}+\frac{p\beta rR_{\rm m}}{N\alpha^{2}b_{\rm m}^{2}}$.
Since these eigenvalues of the fixed point $F_{4}$ for $\tau=0$ satisfies
$\lambda_{1}+\lambda_{2}<0$ and $\lambda_{1}\lambda_{2}>0$, we know that $F_{4}$ is asymptotically stable for $\tau=0$.
Moreover, since $c^{2}<d^{2}$, there is only one pair of purely imaginary solutions and $\lambda=\pm \rm i\omega_{+}$ with $\omega_{+}>0$,
where
\begin{eqnarray}
\omega_{+}^{2}=\frac{1}{2}\left[Q
+\left(Q^{2}+4Np\beta K
(1-K)\frac{\alpha b_{\rm m}}{R_{\rm m}}\right)^{\frac{1}{2}}\right]\,
\label{70} \end{eqnarray}
and $Q=\frac{rp\beta}{\alpha b_{\rm m}}(2r-\frac{3rp\beta}{\alpha b_{\rm m}}-2e_{D})$.
According to reference~\cite{kuang_93} we know that there exists a critical time delay $\tau_{\rm c}=\frac{\theta_{1}}{\omega_{+}}$, where
$\theta_{1}$ satisfies
\begin{equation}
\cos\theta_{1}=-\frac{(ab^{2}-d)\omega_{+}^{2}}{b^{2}\omega_{+}^{2}+d^{2}} \hspace{0.5cm}
%\label{71} \end{equation}
\textrm{and} \hspace{0.5cm}
%\begin{equation}
\sin\theta_{1}=-\frac{ad^{2}\omega_{+}+b\omega_{+}^{3}}{b^{2}\omega_{+}^{2}+d^{2}}\,.
\label{72} \end{equation}
Hence $F_{4}$ is asymptotically stable for $\tau<\tau_{\rm c}$ and unstable for $\tau>\tau_{\rm c}$, and there exists a Hopf bifurcation for $\tau=\tau_{\rm c}$.
Furthermore, we can determine the direction of Hopf bifurcation and the stability of the bifurcation periodic solutions by the analysis mentioned before.

In the case of $r>e_{D}$ and $\alpha b_{\rm m}(1-\frac{e_{D}}{r})> p\beta$,
the equation system has four fixed points which are $F_{0}$, $F_{1}$, $F_{2}$, and $F_{3}$, respectively. The characteristic equation of equation (\ref{2}) at the fixed point $F_{0}$ is
\begin{equation}
\lambda^{2}-(r-e_{D}+p\beta)\lambda+p\beta(r-e_{D})=0.
\label{73} \end{equation}
Since $H<0$, $c^{2}-d^{2}>0$, and the eigenvalues are $\lambda_{1}=r-e_{D}>0$ and $\lambda_{2}=p\beta>0$, $F_{0}$ is unstable for $\tau\geq0$ according to Proposition 1 and reference~\cite{kuang_93}.

The characteristic equation of equation (\ref{2}) at the fixed point $F_{1}$ is
\begin{equation}
\lambda^{2}-(r-e_{D}-p\beta)\lambda-\alpha e_{D} \lambda {\rm e^{-\lambda\tau}}
-p\beta(r-e_{D})-p\beta\alpha e_{D}{\rm e^{-\lambda\tau}} =0.
\label{74} \end{equation}
These eigenvalues of the fixed point $F_{1}$
for $\tau=0$ are $\lambda_{1}=-p\beta<0$ and $\lambda_{2}=r-e_{C}>0$, respectively.
Therefore, $F_{1}$ is unstable for $\tau=0$.
Moreover, for $e_{D}<r<e_{D}(1+\alpha)$, since $c^{2}<d^{2}$, there exist a pair of purely imaginary solutions,
which are $\lambda=\pm \rm i\omega_{+}$ with $\omega_{+}>0$.
We further obtain ${\frac{d(R_{e}\lambda)}{d\tau}}|_{\lambda=\rm i\omega_{+}}>0$, therefore $F_{1}$ remains unstable for $\tau\geq0$~\cite{kuang_93}. While for $r>e_{D}(1+\alpha)$, since $H<0$ and $c^{2}>d^{2}$, there do not exist a pair of purely imaginary solutions and there are no stability switches for any $\tau\geq0$ according to Proposition 1. Therefore, the stability of the fixed point $F_{1}$ for $\tau>0$ is the same with $\tau=0$, and $F_{1}$ remains unstable for $\tau\geq0$~\cite{kuang_93}.

The characteristic equation of equation (\ref{2}) at the fixed point $F_{2}$ is
\begin{equation}
\lambda^{2}-(e_{D}-r+p\beta-\alpha b_{\rm m}+\alpha b_{\rm m}e_{D})\lambda
+(p\beta-\alpha b_{\rm m}+\alpha b_{\rm m}e_{D})(r-e_{D})=0.
\label{76} \end{equation}
Since $H<0$, and $c^{2}-d^{2}>0$, and the eigenvalues are $\lambda_{1}=p\beta-\alpha b_{\rm m}+\alpha b_{\rm m}\frac{e_{D}}{r}<0$ and $\lambda_{2}=e_{D}-r<0$, according to Proposition 1 and reference~\cite{kuang_93} $F_{2}$ is asymptotically stable for $\tau\geq0$.

The characteristic equation of equation (\ref{2}) at the fixed point $F_{3}$ is
\begin{equation}
\lambda^{2}-(L+M)\lambda- \alpha e_{D}\lambda {\rm e^{-\lambda\tau}}
+ML+M\alpha e_{D}{\rm e^{-\lambda\tau}} =0.
\label{75} \end{equation}
Since $H<0$, $c^{2}-d^{2}>0$, and the eigenvalues are $\lambda_{1}=\alpha b_{\rm m}(1-\frac{e_{C}}{r})-p\beta>0$ and $\lambda_{2}=e_{C}-r<0$, $F_{3}$ remains unstable for $\tau\geq0$ according to Proposition 1 and reference~\cite{kuang_93}.

Finally, we note that there exist two special cases of $p\beta=\alpha b_{\rm m}(1-\frac{e_{C}}{r})$ and $p\beta=\alpha b_{\rm m}(1-\frac{e_{D}}{r})$ for rapidly growing resource pool. We now provide theoretical analysis for the equilibrium points in these special cases.

In the first case of  $p\beta=\alpha b_{\rm m}(1-\frac{e_{D}}{r})$,
we then have $K=1$
and $R_{\rm m}-\frac{Nb_{\rm m}}{r}=\frac{p\beta R_{\rm m}}{\alpha b_{\rm m}}$.
As a result, the equation system has four fixed points which are $F_{0}$, $F_{1}$, $F_{2}$,  and  $F_{3}$. The characteristic equation of equation (\ref{2}) at the fixed point $F_{0}$ is
\begin{equation}
\lambda^{2}-[r-e_{D}+p\beta]\lambda+p\beta(r-e_{D})=0.
\label{77} \end{equation}
Since $H<0$, $c^{2}-d^{2}>0$, and the eigenvalues
are $\lambda_{1}=r-e_{D}>0$ and $\lambda_{2}=p\beta>0$, $F_{0}$ is unstable for $\tau\geq0$ according to Proposition 1 and reference~\cite{kuang_93}.

The characteristic equation of equation(\ref{2}) at the fixed point $F_{1}$ is
\begin{equation}
\lambda^{2}-(r-e_{D}-p\beta)\lambda-\alpha e_{D} \lambda {\rm e^{-\lambda\tau}}
-p\beta(r-e_{D})-p\beta\alpha e_{D}{\rm e^{-\lambda\tau}} =0.
\label{78} \end{equation}
These eigenvalues of the fixed point $F_{1}$
for $\tau=0$ are $\lambda_{1}=-p\beta<0$ and $\lambda_{2}=r-e_{C}>0$, respectively.
Therefore, $F_{1}$ is unstable for $\tau=0$.
Moreover, for $e_{D}<r<e_{D}(1+\alpha)$, since $c^{2}<d^{2}$, there exist a pair of purely imaginary solutions which are $\lambda=\pm \rm i\omega_{+}$ with $\omega_{+}>0$. We further obtain ${\frac{d(R_{e}\lambda)}{d\tau}}|_{\lambda=\rm i\omega_{+}}>0$, therefore $F_{1}$ remains unstable for $\tau\geq0$~\cite{kuang_93}. While for $r>e_{D}(1+\alpha)$, since $H<0$ and $c^{2}>d^{2}$, there do not exist a pair of purely imaginary solutions and there are no stability switches for any $\tau\geq0$ according to Proposition 1.
Therefore, the stability of the fixed point $F_{1}$ for $\tau>0$ is the same with $\tau=0$, and $F_{1}$ remains unstable for $\tau\geq0$~\cite{kuang_93}.

The characteristic equation of equation~(\ref{2}) at the fixed point $F_{2}$ is
\begin{equation}
\lambda^{2}-(e_{D}-r+p\beta-\alpha b_{\rm m}+\frac{\alpha b_{\rm m}e_{D}}{r})\lambda
+(p\beta-\alpha b_{\rm m}+\frac{\alpha b_{\rm m}e_{D}}{r})(r-e_{D})=0.
\label{80} \end{equation}
In this situation, we have $a=r-e_{D}>0$, $b=0$, $c=0$, and $d=0$. Since $a=r-e_{D}>0$ and $a^{2}-b^{2}=(r-e_{D})^{2}>0$, $F_{2}$ is stable, but not asymptotically stable~\cite{kuang_93}.

The characteristic equation of equation (\ref{2}) at the fixed point $F_{3}$ is
\begin{equation}
\lambda^{2}-(L+M)\lambda- \alpha e_{D}\lambda {\rm e^{-\lambda\tau}}
+ML+M\alpha e_{D}{\rm e^{-\lambda\tau}} =0.
\label{79} \end{equation}
Since $H<0$, $c^{2}-d^{2}>0$, and the eigenvalues are $\lambda_{1}=\alpha b_{\rm m}(1-\frac{e_{C}}{r})-p\beta>0$ and $\lambda_{2}=e_{C}-r<0$, according to Proposition 1 and reference~\cite{kuang_93} $F_{3}$ remains unstable for $\tau\geq0$.

In the second case of $p\beta=\alpha b_{\rm m}(1-\frac{e_{C}}{r})$, we then have $K=1$
and $R_{\rm m}-\frac{Nb_{\rm m}(1-\alpha)}{r}=\frac{p\beta R_{\rm m}}{\alpha b_{\rm m}}$.
As a result, the equation system has four fixed points, which are $F_{0}$, $F_{1}$, $F_{2}$,  and  $F_{3}$, respectively. The characteristic equation of equation (\ref{2}) at the fixed point $F_{0}$ is
\begin{equation}
\lambda^{2}-(r-e_{D}+p\beta)\lambda+p\beta(r-e_{D})=0.
\label{81} \end{equation}
Since $H<0$, $c^{2}-d^{2}>0$ and the eigenvalues
are $\lambda_{1}=r-e_{D}>0$ and $\lambda_{2}=p\beta>0$, $F_{0}$ is unstable for $\tau\geq0$ according to Proposition 1 and reference~\cite{kuang_93}.

The characteristic equation of equation (\ref{2}) at the fixed point $F_{1}$ is
\begin{equation}
\lambda^{2}-(r-e_{D}-p\beta)\lambda-\alpha e_{D} \lambda {\rm e^{-\lambda\tau}}
-p\beta(r-e_{D})-p\beta\alpha e_{D}{\rm e^{-\lambda\tau}} =0.
\label{82} \end{equation}
These eigenvalues are $\lambda_{1}=-p\beta<0$ and $\lambda_{2}=r-e_{C}>0$, therefore $F_{1}$ is unstable for $\tau=0$. Moreover, for $e_{D}<r<e_{D}(1+\alpha)$, since $c^{2}<d^{2}$, there exist a pair of purely imaginary solutions which are $\lambda=\pm \rm i\omega_{+}$ with $\omega_{+}>0$. We further obtain ${\frac{d(R_{e}\lambda)}{d\tau}}|_{\lambda=\rm i\omega_{+}}>0$, therefore $F_{1}$ remains unstable for $\tau\geq0$ ~\cite{kuang_93}. While for $r>e_{D}(1+\alpha)$, since $H<0$ and $c^{2}>d^{2}$,
according to Proposition 1 there do not exist a pair of purely imaginary solutions and there are no stability switches for any $\tau\geq0$. Therefore, the stability of the fixed point $F_{1}$ for $\tau>0$ is the same with $\tau=0$, and $F_{1}$ remains unstable for $\tau\geq0$~\cite{kuang_93}.

The characteristic equation of equation (\ref{2}) at the fixed point $F_{2}$ is
\begin{equation}
\lambda^{2}-(e_{D}-r+p\beta-\alpha b_{\rm m}+\frac{\alpha b_{\rm m}e_{D}}{r})\lambda
+(p\beta-\alpha b_{\rm m}+\frac{\alpha b_{\rm m}e_{D}}{r})(r-e_{D})=0.
\label{83} \end{equation}
Since $H<0$, $c^{2}-d^{2}>0$, and the eigenvalues
are $\lambda_{1}=p\beta-\alpha b_{\rm m}+\frac{\alpha b_{\rm m}e_{D}}{r}>0$ and $\lambda_{2}=e_{D}-r<0$,  $F_{2}$ is unstable for $\tau\geq0$ according to Proposition 1 and reference~\cite{kuang_93}.

The characteristic equation of equation (\ref{2}) at the fixed point $F_{3}$ is
\begin{equation}
\lambda^{2}-(L+M)\lambda- \alpha e_{D}\lambda {\rm e^{-\lambda\tau}}
+ML+M\alpha e_{D}{\rm e^{-\lambda\tau}} =0.
\label{84} \end{equation}
In this situation, we have $a=r-e_{C}>0$, $b=0$, $c=0$, and $d=0$.
Since $a=r-e_{C}>0$ and $a^{2}-b^{2}=(r-e_{C})(e_{D}(1-3\alpha)-r)>0$, according to reference~\cite{kuang_93} we know that $F_{3}$ is stable, but not asymptotically stable.

\section*{References}
%\bibliography{myreference}

\begin{thebibliography}{60}

\bibitem{ostrom_90}
Ostrom E 1990 {\em Governing the commons: The evolution of institutions for
  collective action\/} (Cambridge: Cambridge university press)

\bibitem{brander_aer98}
Brander J~A and Taylor M~S 1998 {\em Am. Econ. Rev.\/} {\bf 88} 119--38

\bibitem{hauser_n14}
Hauser O~P, Rand D~G, Peysakhovich A and Nowak M~A 2014 {\em Nature\/} {\bf
  511} 220--3

\bibitem{estrela_tee19}
Estrela S, Libby E, Van~Cleve J, D{\'e}barre F, Deforet M, Harcombe W~R,
  Pe{\~n}a J, Brown S~P and Hochberg M~E 2019 {\em Trends Ecol. Evol.\/} {\bf
  34} 6--18

\bibitem{sugiarto_prl17}
Sugiarto H~S, Lansing J~S, Chung N~N, Lai C, Cheong S~A and Chew L~Y 2017 {\em
  Phys. Rev. Lett.\/} {\bf 118} 208301

\bibitem{santos_pnas11}
Santos F~C and Pacheco J~M 2011 {\em Proc. Natl. Acad. Sci. USA\/} {\bf 108}
  10421--5

\bibitem{sanchez_pb13}
Sanchez A and Gore J 2013 {\em PLOS Biol.\/} {\bf 11} e1001547

\bibitem{allen_pb13}
Allen B and Nowak M~A 2013 {\em PLOS Biol.\/} {\bf 11} e1001549

\bibitem{vasconcelos_pnas14}
Vasconcelos V~V, Santos F~C, Pacheco J~M and Levin S~A 2014 {\em Proc. Natl.
  Acad. Sci. USA\/} {\bf 111} 2212--6

\bibitem{pacheco_plrev14}
Pacheco J~M, Vasconcelos V~V and Santos F~C 2014 {\em Phys. Life Rev.\/} {\bf
  11} 573--86

\bibitem{tavoni_ncc14}
Tavoni A and Levin S 2014 {\em Nat. Clim. Change\/} {\bf 4} 1057--63

\bibitem{hilbe_n18}
Hilbe C, {\v{S}}imsa {\v{S}}, Chatterjee K and Nowak M~A 2018 {\em Nature\/}
  {\bf 559} 246--9

\bibitem{su_pnas19}
Su Q, Mcavoy A, Wang L and Nowak M~A 2019 {\em Proc. Natl. Acad. Sci. USA\/}
  {\bf 116} 25398--404

\bibitem{barfuss_pnas20}
Barfuss W, Donges J~F, Vasconcelos V~V, Kurths J and Levin S~A 2020 {\em Proc.
  Natl. Acad. Sci. USA\/} {\bf 117} 12915--22

\bibitem{tavoni_jtb12}
Tavoni A, Ma J~S and Levin S 2012 {\em J. Theor. Biol.\/} {\bf 299} 152--61

\bibitem{lade_te13}
Lade S~J, Tavoni A, Levin S~A and Schl\"{u}ter M 2013 {\em Theor. Ecol.\/} {\bf 6}
  359--72

\bibitem{weitz_pnas16}
Weitz J~S, Eksin C, Paarporn K, Brown S~P and Ratcliff W~C 2016 {\em Proc.
  Natl. Acad. Sci. USA\/} {\bf 113} E7518--25

\bibitem{lee_jh_jtb17}
Lee J~H, Jusup M and Iwasa Y 2017 {\em J. Theor. Biol.\/} {\bf 428} 76--86

\bibitem{chen_xj_pcb18}
Chen X and Szolnoki A 2018 {\em PLOS Comput. Biol.\/} {\bf 14} e1006347

\bibitem{shao_yx_epl19}
Shao Y, Wang X and Fu F 2019 {\em EPL\/} {\bf 126} 40005

\bibitem{hauert_jtb19}
Hauert C, Saade C and McAvoy A 2019 {\em J. Theor. Biol.\/}
  {\bf 462} 347--60

\bibitem{lin_yh_prl19}
Lin Y~H and Weitz J~S 2019 {\em Phys. Rev. Lett.\/} {\bf 122} 148102

\bibitem{wang_x_rspa20}
Wang X, Zheng Z and Fu F 2020 {\em Proc. R. Soc. A\/} {\bf 476} 20190643

\bibitem{tilman_nc20}
Tilman A~R, Plotkin J~B and Akay E 2020 {\em Nat. Commun.\/} {\bf 11} 915

\bibitem{sigmund_n10}
Sigmund K, De~Silva H, Traulsen A and Hauert C 2010 {\em Nature\/} {\bf 466}
  861--3

\bibitem{vasconcelos_ncc13}
Vasconcelos V~V, Santos F~C and Pacheco J~M 2013 {\em Nat. Clim. Change\/} {\bf
  3} 797--801

\bibitem{han_jrsif15}
Han T~A, Pereira L~M and Lenaerts T 2015 {\em J. R. Soc. Interface\/} {\bf 12}
  20141203

\bibitem{perc_pr17}
Perc M, Jordan J~J, Rand D~G, Wang Z, Boccaletti S and Szolnoki A {\em Phsy.
  Rep.\/} {\bf 687} 1--51

\bibitem{kraak_ff11}
Kraak S~B~M 2011 {\em Fish Fish.\/} {\bf 12} 18--33

\bibitem{bauer_prl18}
Bauer M and Frey E 2018 {\em Phys. Rev. Lett.\/} {\bf 121} 268101

\bibitem{santos_nature08}
Santos F~C, Santos M~D and Pacheco J~M 2008 {\em Nature\/} {\bf 454} 213--6

\bibitem{szolnoki_njp16}
Szolnoki A and Perc M 2016 {\em New J. Phys.\/} {\bf 18} 083021

\bibitem{rauch_jrsif17}
Rauch J, Kondev J and Sanchez A 2017 {\em J. R. Soc. Interface\/} {\bf 14}
  20160967

\bibitem{hassard_81}
Hassard B~D, Hassard B, Kazarinoff N~D and Wan Y~W 1981 {\em Theory
  and applications of Hopf bifurcation\/} vol~41 (New York: Cambridge
  University Press)

\bibitem{tsoularis_mb02}
Tsoularis A and Wallace J 2002 {\em Math. Biosci.\/} {\bf 179} 21--55

\bibitem{yang_pnas13}
Yang W, Liu W, Vi{\~n}a A, Tuanmu M~N, He G, Dietz T and Liu J 2013 {\em Proc.
  Natl. Acad. Sci. USA\/} {\bf 110} 10916--21

\bibitem{chen_njp14}
Chen X, Szolnoki A and Perc M 2014 {\em New J. Phys.\/} {\bf 16} 083016

\bibitem{chen_srep15}
Chen X, Sasaki T and Perc M 2015 {\em Sci. Rep.\/} {\bf 5} 17050

\bibitem{liu_mmmas19}
Liu L, Chen X and Szolnoki A 2019 {\em Math. Models Methods App. Sci.\/} {\bf 29} 2127-2149

\bibitem{hofbauer_98}
Hofbauer J and Sigmund K  1998 {\em Evolutionary games and
  population dynamics\/} (Cambridge: Cambridge University Press)

\bibitem{sandholm_2011}
Sandholm W~H 2011 {\em Population Games and Evolutionary Dynamics\/}
  (Cambridge, MA: MIT Press)

\bibitem{tanimoto_2015}
Tanimoto J 2015 {\em Fundamentals of evolutionary game theory and its
  applications\/} (Springer)

\bibitem{harper_entropy16}
Harper M and Fryer D 2016 {\em Entropy\/} {\bf 18} 316

\bibitem{kuang_93}
Kuang Y 1993 {\em Delay differential equations: with applications in
  population dynamics\/} (New York: Academic Press)

\bibitem{gopalsamy_92}
Gopalsamy K 1992 {\em Stability and Oscillations in Delay Differential
  Equations of Population Dynamics\/} (Boston: Kluwer Academic)

\bibitem{sigdel_jtb17}
Sigdel R~P, Anand M and Bauch C~T 2017 {\em J. Theor. Biol.\/} {\bf 432} 132-140

\bibitem{antonioni_pre19}
Antonioni A, Martinez-Vaquero L~A, Mathis C, Peel L and Stella M 2019 {\em Phys. Rev. E\/} {\bf 99} 052311

\bibitem{dobramysl_jpa18}
Dobramysl U, Mobilia M, Pleimling M and T{\"a}uber U~C 2018 {\em J. Phys. A:
  Math. Theor.\/} {\bf 51} 063001

\bibitem{zheng_xd_prl18}
Zheng X~D, Li C, Lessard S and Tao Y 2018 {\em Phys. Rev. Lett.\/} {\bf 120}
  218101

\bibitem{avelino_pre18}
Avelino P~P, Bazeia D, Losano L, Menezes J, De~Oliveira B~F and Santos M~A 2018
  {\em Phys. Rev. E\/} {\bf 97} 032415

\bibitem{intoy_pre15}
Intoy B and Pleimling M 2015 {\em Phys. Rev. E\/} {\bf 91} 052135

\bibitem{perko_01}
Perko L 2001 {\em Differential equations and dynamical systems\/} (New
  York: Springer)

\bibitem{cao_j_ieee07}
Cao J and Xiao M 2007 {\em IEEE Trans. Neur. Net.\/} {\bf 18} 416--30

\end{thebibliography}

\end{document}